\documentclass[twocolumn]{aastex62}

\usepackage{txfonts}

\usepackage{color,bm}

\newcommand{\eqref}[1]{(\ref{#1})}

\shorttitle{Microphysical Modeling of Mineral Clouds in super-Earths}
\shortauthors{Ohno \& Okuzumi}

\begin{document}

\title{Microphysical Modeling of Mineral Clouds in GJ1214 b and GJ436 b: Predicting Upper Limits on the Cloud-Top Height}

\author{Kazumasa Ohno and Satoshi Okuzumi}
\affil{Department of Earth and Planetary Sciences, Tokyo Institute of Technology, Meguro, Tokyo, 152-8551, Japan}

\begin{abstract}
The ubiquity of clouds in the atmospheres of exoplanets, especially of super-Earths, is one of the outstanding issues for transmission spectra survey.
The understanding about the formation process of clouds in super-Earths is necessary to interpret the observed spectra correctly.
In this study, we investigate the vertical distributions of particle size and mass density of mineral clouds in super-Earths using a microphysical model that takes into account the vertical transport and growth of cloud particles in a self-consistent manner.
We demonstrate that the vertical profiles of mineral clouds significantly vary with the concentration of cloud condensation nuclei and atmospheric metallicity. 
We find that the height of the cloud top increases with increasing metallicity as long as the metallicity is lower than a threshold.
If the metallicity is larger than the threshold, the cloud-top height no longer increases appreciably with metallicity because coalescence yields larger particles of higher settling velocities.
We apply our cloud model to GJ1214 b and GJ436 b for which recent transmission observations suggest the presence of high-altitude opaque clouds.
For GJ436 b, we show that KCl particles can ascend high enough to explain the observation.
For GJ1214 b, by contrast, the height of KCl clouds predicted from our model is too low to explain its flat transmission spectrum. 
Clouds made of highly porous KCl particles could explain the observations if the atmosphere is highly metal-rich, and hence the particle microstructure might be a key to interpret the flat spectrum of GJ1214 b.
\end{abstract}


\section{Introduction} \label{sec:intro}
Transmission spectroscopy is one of the powerful approaches to probe the composition of exoplanetary atmospheres \citep[e.g.,][]{Seager&Sasselov00,Brown01}.
Recent observations of the transmission spectra of super-Earths\footnote{In this paper, we refer to super-Earth as a planet larger than Earth but smaller than Neptune in radius. A planet whose size is close to Neptune rather than the Earth is also called a mini-Neptune.} have revealed that some of them might have hydrogen-rich atmospheres \citep{Fraine+14,Tsiaras+16,Southworth+17,Wakeford+17}.
However, it has also been revealed many super-Earths exhibit featureless spectra that imply the presence of high metallicity atmospheres and/or opaque clouds at high altitude \citep[e.g.,][]{Bean+10,Ehrenreich+14,Kreidberg+14a, Knutson+14a,Knutson+14b,Dragomir+15,Stevenson+16}.
Understanding the origin of these high-altitudes clouds is important because they might offer important clues on the composition and structure of the atmosphere beneath.

GJ1214 b and GJ436 b are the typical super-Earths that show featureless transmission spectra \citep[e.g.,][]{Bean+10, Berta+12,Narita+13,Kreidberg+14a,Knutson+14a}.
\citet{Kreidberg+14a} measured the near-infrared transmission spectrum of GJ1214 b using the {\it Hubble Space Telescope} and found that a cloud-free atmosphere cannot explain the featureless spectrum even if a pure steam atmosphere is assumed.
They showed that the presence of an opaque cloud at pressure below ${10}^{-5}~{\rm bar}$ is necessary to explain the observed spectrum. 
\citet{Knutson+14a} measured the transmission spectrum of GJ436 b using the same instrument and found that the planet has a featureless spectrum that can be explained by high-metallicity ($\sim1000\times$ solar) atmosphere and/or an opaque cloud at ${10}^{-3}~{\rm bar}$.

One possible mechanism that can form high-altitude clouds in super-Earths is condensation from vapor to particles followed by upward transport by convection or turbulent diffusion \citep[e.g.,][]{Ackerman&Marley01} as seen in terrestrial water clouds.
In close-in super-Earths where the atmospheric temperature is $500$--$1000~{\rm K}$, minerals such as KCl and ZnS can condense and form clouds \citep[e.g.,][]{Miller-Ricci+12}.
\citet{Morley+13,Morley+15} investigated the vertical distribution of clouds in GJ1214b using the cloud model of \citet{Ackerman&Marley01}.
They found that mineral clouds can ascend to extremely high altitude as suggested from the observation of \citet{Kreidberg+14a} if a sufficiently low settling velocity for cloud particles is assumed.
Since the settling speed generally increases with the size of the particles, the results of \citet{Morley+13,Morley+15} mean that a high-altitude clouds can form if the cloud particles are sufficiently small.
However, because \citet{Morley+13,Morley+15} parameterized the ratio of the settling velocity to upward velocity as a free parameter, it is still unclear whether the assumed particle size is realistic.
\citet{Morley+17} applied the same model to GJ436 b, and found that a very thick cloud is not favored because such cloud would obscure the molecular lines seen in the observed emission spectrum \citep[e.g.,][]{Stevenson+10}. 
\citet{Charnay+15,Charnay+15b} investigated the global cloud distribution in GJ1214b using a 3D global circulation model (GCM) together with a simple tracer model developed by \citet{Parmentier+13}.
They showed that the large-scale atmospheric circulation driven by the intense day-night heating contrast can loft cloud particles to altitude high enough to obscure the spectral feature if the atmospheric metallicity is higher than $>100\times$ solar and the particle radius is $\sim 0.5~{\rm \mu m}$.
However, the particle size is a free parameter in their studies.

Another candidate for the origin of the flat spectra is organic haze formed through the UV photolysis of carbon-bearing species in the upper atmosphere \citep{Miller-Ricci+12}.
\citet{Morley+13,Morley+15} suggest that photochemical haze can explain the flat spectrum of GJ1214 b if the haze particles are small and if their production rate is high. 
Recently, \citet{Kawashima&Ikoma18} investigated the vertical profiles of haze using both photochemical calculation and particle growth model, and found that the flat transmission spectra would be explained if the haze production rate per unit Ly $\alpha$ intensity is considerably higher than would be expected from Titan's haze.
However, it is yet to be explained why the haze production rate per unit UV irradiation would be so high.

As introduced above, the cloud properties, especially the cloud particle size, for super-Earths are still poorly understood.
In this study, we investigate the vertical structure of mineral clouds in GJ1214 b and GJ436 b to understand how the particle size and number density vary with atmospheric parameters, including the atmospheric metallicity.
We apply a 1D cloud model that takes into account the vertical transport, gravitational settling, condensation, and collisional growth of cloud particles in a self-consistent manner.
The structure of this paper is as follows.
In Section \ref{sec:method}, we describe the basic equations and numerical setting.
In Section \ref{sec:result}, we show the results of calculations and interpretation of the microphysical processed controlling the cloud particle size. 
In Section \ref{sec:observation}, we compare the cloud-top height predicted from our model with those inferred from the observations of GJ1214 b and GJ436 b to examine if the flat spectra of these super-Earths are caused by mineral clouds. 
In Section \ref{sec:discussion}, we mainly discuss how size distribution and particle porosity affect the height of cloud top.
Our conclusions are presented in Section~\ref{sec:conclusions}.

\section{Method} \label{sec:method}
\subsection{Outline}
We extend the microphysical model originally developed by \citet{Ohno&Okuzumi17} to predict the vertical distributions of a cloud in the atmosphere.
The cloud model of \citet{Ohno&Okuzumi17} adopts a 1D Eulerian framework, and provides the vertical distributions of number ($n_{\rm c}$) and mass ($\rho_{\rm c}$) densities of cloud particles by taking into account the vertical transport of cloud particles due to the updraft motion and gravitational settling, and the particle growth via condensation and coalescence (see Section \ref{sec:transport} and \ref{sec:microphysics}).
In this study, we take into account the vertical transport of cloud particles via eddy diffusion \citep[e.g.,][]{Ackerman&Marley01}. 

Following previous studies \citet{Charnay+15,Morley+13,Morley+15}, we consider the clouds composed of solid KCl particles formed through the condensation of KCl vapor.
The initial cloud particles are assumed to form at the cloud base through the condensation of vapor onto the small nuclei that already exist in the atmosphere, the process so called heterogeneous nucleation.
On the Earth, such small nuclei, called the cloud condensation nuclei (CCNs), include sea salt, volcano ash, and dust from the land \citep{Rogers&Yau89}.
The amount of CCNs on exoplanets is still highly uncertain as well as is their composition, and therefore we take the number density of CCNs as a free parameter.
The height of the cloud base is determined from the comparison between the atmospheric temperature and condensation temperature (see Section \ref{sec:PT}). 
The condensation temperature is defined as the temperature at which the partial pressure of a volatile is equal to its saturation vapor pressure.

Following \citet{Ohno&Okuzumi17}, we assume that the cloud particles have the characteristic radius $r_{\rm c}$ and corresponding mass $m_{\rm c}=(4\pi/3)\rho_{\rm int}r_{\rm c}^3$, where $\rho_{\rm int}$ is the internal density of the particles.
The internal density can vary significantly if the particles grow into porous aggregates \citep{Kataoka+13}.
In this study we simply assume $\rho_{\rm int}=\rho_{\rm p}$, where $\rho_{\rm p}$ is the material density of the condensate, but we will discuss the influences of varying the internal density in Section \ref{sec:porosity}.
Assuming the mass distribution is narrowly peaked at $m\approx m_{\rm c}$, the number and mass densities are related by $\rho_{\rm c}=m_{\rm c}n_{\rm c}$.
Such frameworks are called the double-moment bulk schemes in meteorology \citep[e.g.,][]{Ziegler85, Ferrier94} and the characteristic size method in planetary formation community \citep[e.g.,][]{Birnstiel+12,Ormel14,Sato+16}. 
This method allows us to derive the physical understanding from calculations more clearly, and to perform the calculations with much little computational time compared to spectral bin schemes \citep[e.g.,][]{Brauer+08} that solve the evolution of the full size distribution.

We investigate the influences of atmospheric metallicity on the vertical profiles of clouds in super-Earths.
In this paper, the atmospheric metallicity refers to the ratio of atmospheric heavy element abundance to that of the solar atmosphere, i.e., $(N_{\rm Z}/(N_{\rm H}+N_{\rm He}))/(N_{\rm Z}/(N_{\rm H}+N_{\rm He}))_{\rm solar}$.
Recent theoretical studies suggested that the atmospheres of super-Earths potentially have the metallicities higher than solar, and even higher than $100\times$ solar, depending on the properties of the building blocks of planets \citep{Fortney+13,Venturini+16}.
Also the interior modeling showed that GJ1214 b might have a steam atmosphere mainly composed of water vapor \citep{Rogers&Seager10,Valencia+13}.
Therefore, we take the atmospheric metallicity as a free parameter widely ranging from the metallicity of $1\times$ solar to water vapor atmosphere.
The metallicity difference provides the different pressure-temperature structure, total cloud mass, and eddy diffusion coefficient.

\subsection{Construction of Vertical Structure}\label{sec:PT}
\begin{figure}
\includegraphics[clip,width=\hsize]{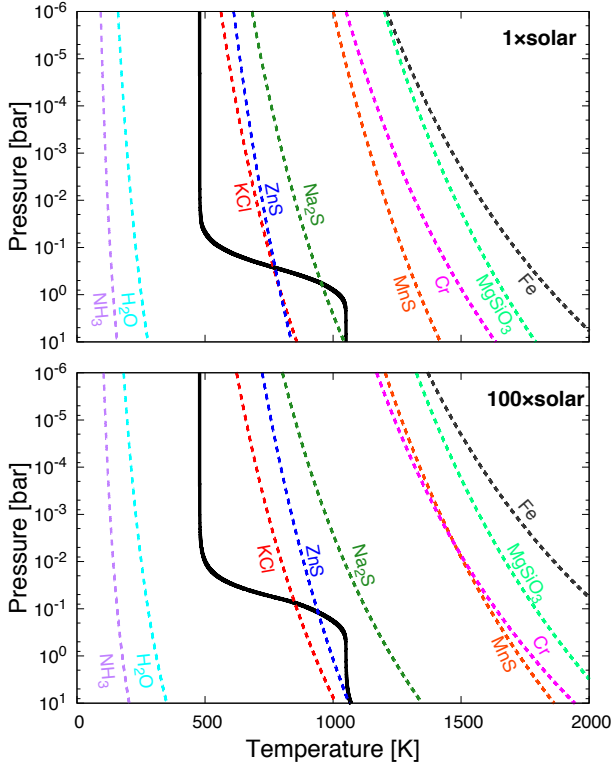}
\caption{P-T profiles of GJ1214 b and the vapor pressure curves for the metallicity of $1\times$ solar (top) and $100\times$ solar (bottom) abundance, respectively.
The vapor pressures used in the figures are taken from \citet{Rogers&Yau89,Ackerman&Marley01,Morley+12}.
The solid black lines are the P-T structure assuming the heat redistribution around the entire planet ($f=1/4$).}
\label{fig:PT}
\end{figure}
To determine the location of the cloud base, we construct the pressure-temperature structure using the analytical model of radiative atmosphere described by \citet{Guillot10} under the assumption of hydrostatic equilibrium.
\citet{Guillot10} derived the analytical solution of global mean thermal profiles that gives good agreement with the predictions from sophisticated simulations.
The stellar effective temperature, radii, semi-major axis, and planetary radii of GJ1214 b and GJ436 b are taken from the Exoplanet.eu catalog \footnote{http://exoplanet.eu}.
Following \citet{Guillot10}, the temperature in each atmospheric layer is given by
\begin{eqnarray}
\nonumber
T^{\rm 4} &=& \frac{3T_{\rm int}^{\rm 4}}{4}\left[ \frac{2}{3}+\tau \right]+\frac{3T_{\rm irr}^{\rm 4}}{4}f\\
&&\times \left[ \frac{2}{3}+\frac{1}{\gamma \sqrt{3}}+\left( \frac{\gamma}{\sqrt{3}} - \frac{1}{\gamma \sqrt{3}}\exp{(-\gamma \sqrt{3} \tau)} \right) \right],
\end{eqnarray}
where $\tau$ is the vertical infrared optical depth $\tau$ is given by
\begin{equation}
\tau(z)=\int_{z}^{\infty}\rho_{\rm g}\kappa_{\rm th}dz',
\end{equation}
where $\kappa_{\rm th}$ is the atmospheric infrared opacity.
The $f=1/4$ is the heat redistribution factor under the assumption of the radiation redistributed around the entire planet, $T_{\rm int}$ is the intrinsic effective temperature, $T_{\rm irr}$ is the irradiation effective temperature, and the $\gamma=\kappa_{\rm v}/\kappa_{\rm th}$ is the ratio of the visible to infrared opacities, respectively.
For GJ1214 b, we take $T_{\rm int}=60~{\rm K}$ \citep{Rogers&Seager10} and $\gamma=0.038$ so that reproduces the P-T structure predicted by radiative transfer models of \citet{Miller-Ricci&Fortney10}.
For GJ436 b, we take $T_{\rm int}=300~{\rm K}$ \citep{Morley+17} and $\gamma=0.05$ that is in a good agreement with the retrieved P-T structure \citep{Miguel+15}.

We calculate $\tau$ using the fitting formula of Rosseland mean opacity of a cloud-flee atmosphere described by \citet{Freedman+14}.
This fitting formula is a function of atmospheric metallicity, pressure, and temperature, and valid for $P={10}^{-6}$--$3\times {10}^{2}~{\rm bar}$ and $T=75$--$4000~{\rm K}$.
Although the opacity table for higher metallicity ($>50\times$ solar) is not available so far, the fitting formula can provide the qualitative results for such high metallicity atmospheres.
For water vapor atmosphere, we use the opacity of $50\times$ solar metallicity that yields the similar P-T structure to that for a water vapor \citep{Miller-Ricci&Fortney10} for simplify.
We also neglect the opacity of cloud particles that might change the location of cloud base, but we plan to investigate this impacts in future study.

Figure \ref{fig:PT} shows the vertical P-T structures of GJ1214 b for $1\times$ and $100\times$ solar metallicity and the condensation temperature at each atmospheric layer.
Here we predict the condensation temperature for each volatile using the saturation vapor pressure described in \citet{Rogers&Yau89,Ackerman&Marley01,Morley+12}.
The vapor species has a solid phase if the atmospheric temperature is lower than its condensation temperature.
Therefore, for each volatile species, the cloud base is expected to be placed at the location where the P-T curve intersects the curve of condensation temperature of the species.
Figure \ref{fig:PT} indicates that the KCl, ZnS, and $\rm {{Na}_{2}S}$ are condensible for $1\times$ solar metallicity case, and KCl and ZnS are condensible for $100\times$ solar metallicity.
Since the abundance of KCl vapor is higher than that of ZnS vapor for solar like atmosphere \citep{Morley+12}, we focus on the mineral clouds of KCl in this study.
The cloud base for KCl is placed at $\sim0.4~{\rm bar}$ for $1\times$ solar metallicity, $\sim0.1~{\rm bar}$ for $10\times$ solar metallicity, and $\sim0.07~{\rm bar}$ for $100\times$ solar metallicity, respectively, which is in good agreement with the prediction of previous studies \citep{Miller-Ricci+12,Morley+13,Charnay+15}.

\subsection{Transport Equations}\label{sec:transport}
We calculate the vertical distributions of the number and mass densities of cloud particles by taking into account their growth and vertical transport.
The microphysics of cloud formation is complex \citep[see e.g.,][]{Rossow78,Rogers&Yau89,Pruppacher&Klett97,Seinfeld&Pandis06}.
However, \citet{Ohno&Okuzumi17} showed that inclusion of condensation and collisional growth is enough to approximately reproduce the observations of terrestrial water clouds and Jovian ammonia clouds.
Therefore, we take into account the condensation and collisional growth in this study.

Following \citet{Charnay+15}, we consider the clouds formed through the large scale atmospheric motion driven by the intense day-night heating contrast.
Previous studies showed that the global averaged distributions of such clouds can be approximately reproduced by a 1D advection-diffusion model with an empirical parameterization of the eddy diffusion coefficient $K_{\rm z}$ \citep{Parmentier+13,Charnay+15}.
Hence, the master equations used here are constructed by adding the source terms expressing particle growth to the 1D advection-diffusion model:
\begin{equation}\label{eq:master1}
\frac{\partial n_{\rm c}}{\partial t}=\frac{\partial}{\partial z}
\left[  n_{\rm g}K_{\rm z}\frac{\partial}{\partial z}\left( \frac{n_{\rm c}}{n_{\rm g}}\right)+v_{\rm t}(r)n_{\rm c} \right]
-\left| \frac{\partial n_{\rm c}}{\partial t} \right|_{\rm coll},
\end{equation}
\begin{equation}\label{eq:master2}
\frac{\partial \rho_{\rm c}}{\partial t}=\frac{\partial}{\partial z}
\left[  \rho_{\rm g}K_{\rm z}\frac{\partial}{\partial z}\left( \frac{\rho_{\rm c}}{\rho_{\rm g}}\right)+v_{\rm t}(r)\rho_{\rm c} \right]+\left( \frac{\partial \rho_{\rm c}}{\partial t}\right)_{\rm cond},
\end{equation}
\begin{equation}\label{eq:master3}
\frac{\partial \rho_{\rm v}}{\partial t}=\frac{\partial}{\partial z}
\left[ \rho_{\rm g}K_{\rm z}\frac{\partial}{\partial z}\left( \frac{\rho_{\rm v}}{\rho_{\rm g}}\right) \right]-\left( \frac{\partial \rho_{\rm c}}{\partial t}\right)_{\rm cond},
\end{equation}
where $v_{\rm t}$ is the terminal velocity of cloud particles, $K_{\rm z}$ is the eddy diffusion coefficient, and $\rho_{\rm v}$ is the vapor mass density.
The terminal velocity depends on the particle size and atmospheric density as introduced in Section \ref{sec:gasdrag}.
Each source term, introduced in Section \ref{sec:microphysics}, expresses the particle growth via condensation and collision of each particle.
Without these terms, the Equations \eqref{eq:master1}--\eqref{eq:master2} are reduced to the 1D transport model for fixed size particles used by \citet{Parmentier+13} and \citet{Charnay+15}.

The eddy diffusion coefficient $K_{\rm z}$ represents the strength of effective vertical mixing for cloud particles.
In this study, we adopt the empirical formula of $K_{\rm z}$ proposed by \citet{Charnay+15}, 
\begin{equation}\label{eq:Kz}
K_{\rm z}=K_{\rm 0}\left( \frac{P}{P_{\rm 0}}\right)^{-2/5},
\end{equation}
where $K_{\rm 0}$ is the value of $K_{\rm z}$ at a reference pressure $P_{\rm 0}$.
\citet{Charnay+15} derived this formula from 3D GCM simulations that takes into account the transport of fixed size particles.
Since they suggested that $K_{\rm z}$ is almost independent of particle size \citep[see the figure 14 of][]{Charnay+15}, we use Equation \eqref{eq:Kz} for all range of particle size in our calculations.
The exponent of $-2/5$ is similar to the $K_{\rm z}\propto P^{-1/3}$ predicted by mixing theory \citep{Ackerman&Marley01} and $K_{\rm z}\propto P^{-1/2}$ predicted by other GCM simulations for hot Jupiter \citep{Parmentier+13}.
According to \citet{Charnay+15}, we choose the reference pressure of $P_{\rm 0}=1~{\rm bar}$ and take the values of $K_{\rm 0}$ 
as summarized in Tables \ref{table:1} and \ref{table:2}.
For GJ1214 b, we use the values of $K_{\rm 0}$ derived from the power-law fitting to the GCM data \citep[see Figure 14 of][]{Charnay+15}, which are metallicity-depenent.
For GJ436 b, we adopt $K_{\rm 0}=2.5\times{10}^{3}~{\rm m^2~s^{-1}}$ indenependently of the metallicity.
The adopted value is similar to the value $K_{\rm z}\sim {10}^2$--${10}^3~{\rm m^2~s^{-1}}$ suggested by \citet{Madhududhan&Seager11} to explain the disequilibrium chemistry seen in GJ436 b's emission spectra.
Our adopted value is two orders of magnitude lower than the earlier prediction by \citet{Lewis+10} based on the rms velocity from 3D GCM calculations. However, \citet{Parmentier+13} and \citet{Charnay+15} recently pointed out that the values of the eddy diffusion coefficient predicted in this way tend to be one or two-orders of magnitude higher than those directly determined from vertical particle distribution. If we take this into account, our choice of $K_0$ is consistent with the GCM results by \citet{Lewis+10}.

\begin{table}[t]
  \centering
   \caption{Model parameters for GJ1214 b }
  \begin{tabular}{c|cccc}
     metallicity & $H$ (km) & $q_{\rm KCl}$ (mol/mol) & $K_{\rm 0}~({\rm m^{2}~s^{-1}})$ & $\Delta z$~(km) \\ \hline \hline
     1$\times$solar &  190 &2.54$\times {10}^{-7}$ & $7.0\times {10}^2$&20\\ 
    10$\times$solar & 180 & 2.52$\times {10}^{-6}$ & $2.8\times {10}^3$&20\\ 
        100$\times$solar & 103 & 2.32$\times {10}^{-5}$ & $3.0\times {10}^3 $&10\\
     Steam & 25  & 2.61$\times {10}^{-4}$ & $3.0\times {10}^2 $ & 5 
  \end{tabular}
  \label{table:1}
  
     \caption{Model parameters for GJ436 b }
  \begin{tabular}{c|cccc}
     metallicity & $H$ (km) & $q_{\rm KCl}$ (mol/mol) & $K_{\rm 0}~({\rm m^{2}~s^{-1}})$ & $\Delta z$~(km) \\ \hline \hline
     1$\times$solar &  169 &2.54$\times {10}^{-7}$ & $2.5\times {10}^3$&20\\ 
    10$\times$solar & 159 & 2.52$\times {10}^{-6}$ & $2.5\times {10}^3$&20\\ 
        100$\times$solar & 102 & 2.32$\times {10}^{-5}$ & $2.5\times {10}^3 $&10\\
     1000$\times$solar & 22  & 2.61$\times {10}^{-4}$ & $2.5\times {10}^3 $ & 5
  \end{tabular}
  \label{table:2}
\end{table}

\subsection{Expression of the Terminal Velocity}\label{sec:gasdrag}
\begin{figure}
\includegraphics[clip,width=\hsize]{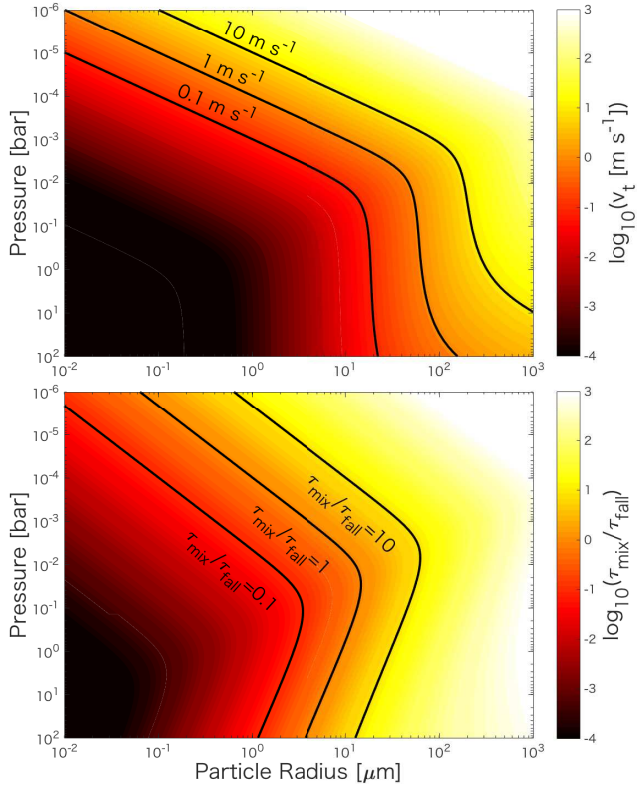}
\caption{Terminal velocity of compact KCl particles (colorscale, top panel) and the ratio of the mixing timescale to the falling timescale (colorscale, bottom panel). The horizontal axis shows particle radius and the vertical axis shows atmospheric pressure, respectively. Each black contour shows the pressure and particle radius corresponding to $v_{\rm t}=0.1$, $1$, and $10~{\rm m~s^{-1}}$ for the top panel, and $\tau_{\rm mix}/\tau_{\rm fall}=0.1$, $1$, and $10$ for the bottom panel, respectively. Here we assume $1\times$ solar metallicity, $K_{\rm 0}={10}^3~{\rm m^{2}~s^{-2}}$, and isothermal ($T=500~{\rm K}$) atmosphere.}
\label{fig:vset}
\end{figure}

A terminal velocity $v_{\rm t}$ is determined by the balance between gravitational force and gas frictional force.
The gas frictional force depends on the behavior of the gas flow around the settling particles, and varies with the particle size, settling velocity, and the mean free path of gas particles \citep[e.g.,][]{Rossow78, Woitke&Helling03}.
In this study, we adopt the following formula of the terminal velocity,
\begin{equation}\label{eq:vt}
    v_{\rm t}(r_{\rm c}) = \frac{2\beta gr_{\rm c}^2 \rho_{\rm p}}{9\eta}\left[ 1+\left(\frac{0.45gr_{\rm c}^3\rho_{\rm g}\rho_{\rm p}}{54\eta^2}\right)^{2/5} \right]^{-5/4},
\end{equation}
where $\eta$ is the dynamic viscosity of the atmosphere and $\beta$ is the {\it slip correction factor}. $\beta$ accounts for the transition of gas drag behavior from viscous flow (Stokes's law) to free molecular flow (Epstein's law) around the particle, given by \citep{Davies45}
\begin{equation}\label{eq:slip}
\beta = 1+{\rm Kn}_{\rm g}[1.257+0.4\exp{(-1.1/{\rm Kn}_{\rm g})}],
\end{equation}
where ${\rm Kn}_{g}=l/r_{\rm c}$ is the gas Knudsen number and $l$ is the gas mean free path given in Appendix \ref{appendix:1}.
Equation \eqref{eq:vt} without $\beta$ is same as the Equation (23) in \citet{Ohno&Okuzumi17} that asymptotically reaches the Stokes's law for a laminar flow limit, Newton's law for a turbulent flow limit, and well reproduces the intermediate regime predicted by experiment \citep[see the Figure 7 in][]{Ohno&Okuzumi17}.
Top panel of Figure \ref{fig:vset} shows the terminal velocity as a function of particle size and atmospheric pressure.
Figure \ref{fig:vset} shows the terminal velocity increases with height in the upper atmosphere because of the Epstein's law arisen from the low atmospheric density.

We also show the ratio of the mixing timescale $\tau_{\rm mix}$ to the falling timescale $\tau_{\rm fall}$ in the bottom panel of Figure \ref{fig:vset}.
Each timescale is defined as
\begin{equation}\label{eq:t_mix}
\tau_{\rm mix}=\frac{H^2}{K_{\rm z}}
\end{equation}
and
\begin{equation}
\tau_{\rm fall}=\frac{H}{v_{\rm t}},
\end{equation}
where $H=k_{\rm B}T/m_{\rm g}$ is the pressure scale height, respectively.
Here we assumed $T=500~{\rm K}$, $K_{\rm 0}={10}^{3}~{\rm m^2~s^{-1}}$, and a solar composition atmosphere.
Cloud particles ascend if $\tau_{\rm mix}\ll \tau_{\rm fall}$, and fall if $\tau_{\rm mix}\gg \tau_{\rm fall}$.
Figure \ref{fig:vset} indicates that the cloud particles are required to maintain their size $r_{\rm c}\la0.5~{\rm \mu m}$ to ascend above ${10}^{-3}~{\rm bar}$ suggested for GJ436b \citep{Knutson+14a}, and $r\la0.05~{\rm \mu m}$ to ascend above ${10}^{-5}~{ \rm bar}$ under the assumed parameters.

\subsection{Microphysics of Particle Growth}\label{sec:microphysics}
The cloud particles ascend from the cloud base while growing through condensation and collision with each other.
Condensation dominates the growth of small particles due to the relatively short timescale.
The growth rate of $\rho_{\rm c}$ via condensation depends on the behavior of vapor molecule motion, and is expressed by \citep{Rogers&Yau89,Woitke&Helling03}
\begin{eqnarray}\label{eq:drhoc/dt}
\left(\frac{\partial \rho_{\rm c}}{\partial t} \right)_{\rm cond}  =  &&4\pi r_{\rm c}^2n_{\rm c}(\rho_{\rm v}-\rho_{\rm s})\times \\
\nonumber
&& {\rm min}\left[C_{\rm re}, \frac{D}{r_{\rm c}}\left(1+\left(\frac{m_{\rm v}L}{k_{\rm B}T}-1  \right)\frac{LD\rho_{\rm s}}{KT} \right)^{-1}\right],
\end{eqnarray}
where $\rho_{\rm v}$ is the vapor mass density, $\rho_{\rm s}$ is the saturation vapor density, $C_{\rm re}=\sqrt{k_{\rm B}T/2\pi m_{\rm v}}$ is the relative velocity of vapor molecules, $m_{\rm v}$ is the mass of the vapor molecules, $L$ is the specific latent heat of condensation, and $D$ is the molecular diffusion coefficient of vapor in ambient air, respectively.
The first formula in the bracket corresponds to the free molecular flow regime \citep[][]{Woitke&Helling03} in which the vapor molecules are freely impinging onto the particles. The second formula corresponds to the diffusive regime \citep{Rogers&Yau89} in which the vapor molecules behave as continuum.

Collisional growth is induced by the relative velocity arisen from both gravitational settling and Brownian motion of particles.
In this paper, we refer the collisional growth by gravitational settling as {\it coalescence} and that by Brownian motion as {\it coagulation}.
Then the decrease in number density via collisional growth is expressed by
\begin{equation}
\left| \frac{\partial n_{\rm c}}{\partial t} \right|_{\rm coll}= \left| \frac{\partial n_{\rm c}}{\partial t} \right|_{\rm coag}+\left| \frac{\partial n_{\rm c}}{\partial t} \right|_{\rm coal},
\end{equation}
where $|\partial n_{\rm c}/\partial t|_{\rm coag}$ is the decrease in number density for coagulation and $|\partial n_{\rm c}/\partial t|_{\rm coal}$ is that for coalescence.
The expression of $|\partial n_{\rm c}/\partial t|_{\rm coag}$ depends on particle Knudsen number $\rm{Kn}_{p}$ defined as
\begin{equation}
{\rm {Kn}_{p}}=\frac{\beta}{6\eta r_{\rm c}^2}\sqrt{ \frac{m_{\rm c}k_{\rm B}T}{2\pi} },
\end{equation}
The Brownian motion of particles is diffusive for $\rm{Kn}_p\ll1$ and ballistic for $\rm{Kn}_p\gg1$.
The rate of decrease of particle number density via coagulation is given by \citep{Seinfeld&Pandis06}
\begin{equation}\label{eq:dcoag}
    \left| \frac{\partial n_{\rm c}}{\partial t} \right|_{\rm coag}  =  
\left\{
\begin{array}{ll}
{\displaystyle 8\sqrt{\frac{\pi k_{\rm B}T}{m_{\rm c}}}r_{\rm c}^2n_{\rm c}^2} & \text{(${\rm Kn_p}>1/\sqrt{2}$)} \\[1.5ex]
{\displaystyle \frac{4k_{\rm B}T\beta}{3\eta}n_{\rm c}^2}  & \text{(${\rm Kn_p}<1/\sqrt{2}$)},
         \end{array}
\right.
\end{equation}
The transition takes pace at $r_{\rm c}\approx 0.07~{\rm \mu m}$ under the assumptions of $T=1000~{\rm K}$, $P=0.1~{\rm bar}$, and $m_{\rm g}=2~{\rm amu}$, which are equivalent to the parameters for the cloud base.

For the coalescence growth, the rate of decrease of number density $|\partial n_{\rm c}/\partial t|_{\rm coal}$ is given by \citep{Rossow78}
\begin{equation}\label{eq:dcoal}
 \left| \frac{\partial n_{\rm c}}{\partial t} \right|_{\rm coal} \approx 2\pi r_{\rm c}^2 n_{\rm c}^2 \Delta vE,
\end{equation}
where $\Delta v$ is the relative velocity induced by the gravitational settling, and $E$ is the collection efficiency defined as the ratio of the effective collisional cross section to the geometric cross section \citep[e.g.,][]{Pruppacher&Klett97}.
For the relative velocity, \citet{Sato+16} and \citet{Krijt+16} showed that the characteristic size approach with $\Delta v=0.5v_{\rm t}(r)$ is in good agreement with the results of spectral bin schemes, and therefore we assume $\Delta v=0.5v_{\rm t}(r_{\rm c})$.
The collection efficiency $E$ accounts for the effect of the gas flow around the particle moving relative to the background gas, and is expressed in terms of Stokes number
\begin{equation}
{\rm Stk} = \frac{v_{\rm t}(r_{\rm c})\Delta v}{gr_{\rm c}},
\end{equation}
which is defined as the ratio of the stopping time $=v_{\rm t}(r_{\rm c})/g$ to the crossing time $\sim r_{\rm c}/\Delta v$.
When ${\rm Stk}\ll1$, the particles is strongly coupled to the gas flow around the another particles, and hence $E$ behaves as $E\approx 0$ \citep{Rossow78}.
We evaluate $E$ using a smoother analytic function of \citet{Guillot+14} given by
\begin{equation}\label{eq:collection}
E={\rm max}[0,1-0.42{\rm Stk}^{-0.75}],
\end{equation}
which vanishes at ${\rm Stk}\la 0.3$ and approaches unity at ${\rm Stk}\gg 1$.
If ${\rm Kn}_{\rm g}>1$, we assumed $E=1$ because the influence of the gas on the particle trajectory should be weak in that region \citep{Rossow78}.

\subsection{Numerical Procedure} \label{sec:setting}
We numerically solve the Equations \eqref{eq:master1}--\eqref{eq:master3} until the system reaches to the steady-state profiles.
The initial number density of the cloud particles at the cloud base is parameterized by the CCN number density $n_{\rm CCN}$.
We take the $n_{\rm CCN}$ as a free parameter widely ranging as ${10}^{6}$--${10}^{15}~{\rm m^{-3}}$.
Since the composition of the CCNs in close-in super-Earths is unknown, we assume the bulk density of the CCN as that of KCl.
This assumption does not affect the calculated cloud vertical profiles as long as the mass fraction of the CCNs in the cloud particles is very small.
Therefore, we choose the upper limit of $n_{\rm CCN}$ so that the total mass of CCNs does not exceed that of KCl vapor at the cloud base.
We set the radii of CCNs as $r_{\rm CCN}=0.001~{\rm \mu m}$, and then $n_{\rm CCN}\approx{10}^{15}~{\rm m^{-3}}$ corresponds to the upper limit for our calculations.

We choose the flux of a lower boundary condition so that $n_{\rm c}/n_{\rm g}$, $\rho_{\rm c}/\rho_{\rm g}$, and $\rho_{\rm v}/\rho_{\rm g}$ keep the values of the cloud base.
We adopt the zero-flux boundary condition at the top of the computational domain which is located at $P={10}^{-8}~{\rm bar}$.
The vertical coordinate $z$ is discretized into linearly spaced bins.
We use the different grid width for different atmospheric metallicity as summarized in Table \ref{table:1}.
The time increment $\Delta t$ is chosen at every time step so that the fractional decreases in $n$, $\rho_{\rm c}$, and $\rho_{\rm v}$ do not exceed $0.5$, i.e., $\Delta t \leq -0.5\times{\rm min}[(\partial ~{\rm ln}~n/\partial t)^{-1},(\partial ~{\rm ln}~\rho_{\rm c}/\partial t)^{-1},(\partial ~{\rm ln}~\rho_{\rm v}/\partial t)^{-1}]$.
However, this expression yields very small $\Delta t$ because the time increment determined by condensation is much shorter than that for collisional growth and vertical transport.
To avoid it, we adjust the time increment as $\Delta t \leq -0.5\times (\partial ~{\rm ln}~n/\partial t)^{-1}$ if $(\partial ~{\rm ln}~\rho_{\rm c}/\partial t)^{-1}<0.1\times(\partial ~{\rm ln}~n/\partial t)^{-1}$.
In this case, we convert the all excess/lack of vapor from saturation value into cloud particles.

We calculate the mean molecular weight of the atmosphere assuming hydrogen-helium-water mixture in accordance with \citet{Fortney+13}.
Elemental abundances are taken from \citet{Lodders03}.
The mixing ratio of KCl vapor $q_{\rm KCl}$ below the cloud base is calculated assuming the number of KCl molecules is equal to that of K.
For the steam atmosphere and the metallicity of $1000\times$ solar, we evaluate the mean molecular weight as that of water, and $q_{\rm KCl}$ as a ratio of K to O because the atmosphere is dominated by water rather than hydrogen for extremely metal-enriched cases.
Appendix \ref{appendix:1} summarizes the evaluation of other physical parameters (e.g., viscosity) required for our calculations.
Table \ref{table:1} and \ref{table:2} show the $q_{\rm KCl}$, $K_{\rm 0}$, and $H$ at the upper isothermal region for GJ1214 b and GJ436 b, respectively.

\section{Results}\label{sec:result}
\subsection{Vertical Distribution of the Particle Size and Mass Density}\label{sec:size}
\begin{figure*}[t]
\includegraphics[clip,width=0.98\hsize]{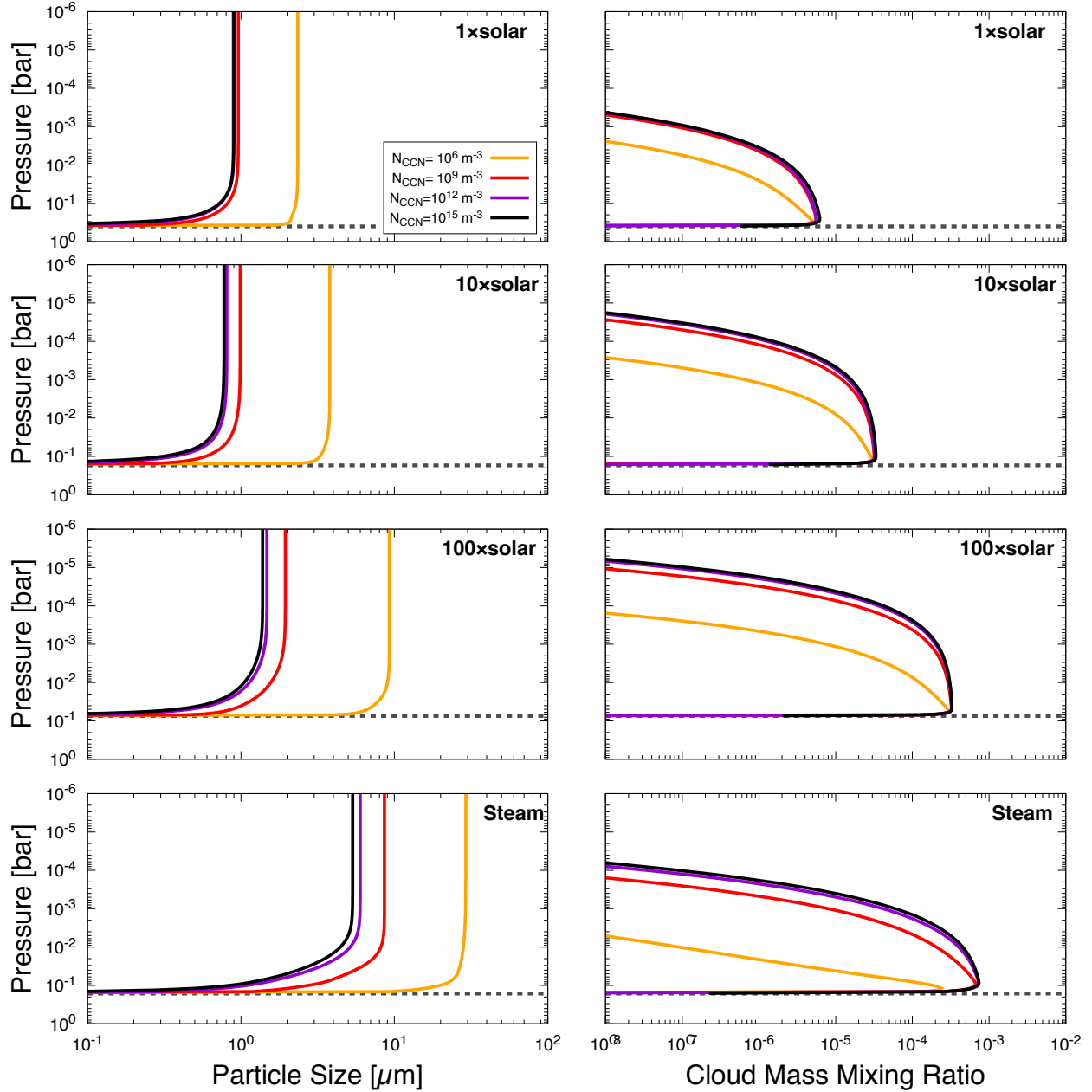}
\caption{Vertical structure of the KCl cloud for different atmospheric metallicity models. The left and right columns show the vertical distributions of the particle radius and mass mixing ratio, respectively, for different values of the CCN number density $n_{\rm CCN}$. 
Each row, from top to bottom, is for atmospheric metallicities of $1\times$, $10\times$, $100\times$ solar, and steam atmosphere, respectively.
The orange, red, purple, and black lines show the results for $n_{\rm CCN}={10}^{6}$, ${10}^{9}$, ${10}^{12}$, and ${10}^{15}~{\rm m^{-3}}$, respectively.
The gray dotted lines indicate the cloud base.
}
\label{fig:size}
\end{figure*}
\begin{figure*}[t]
\includegraphics[clip,width=\hsize]{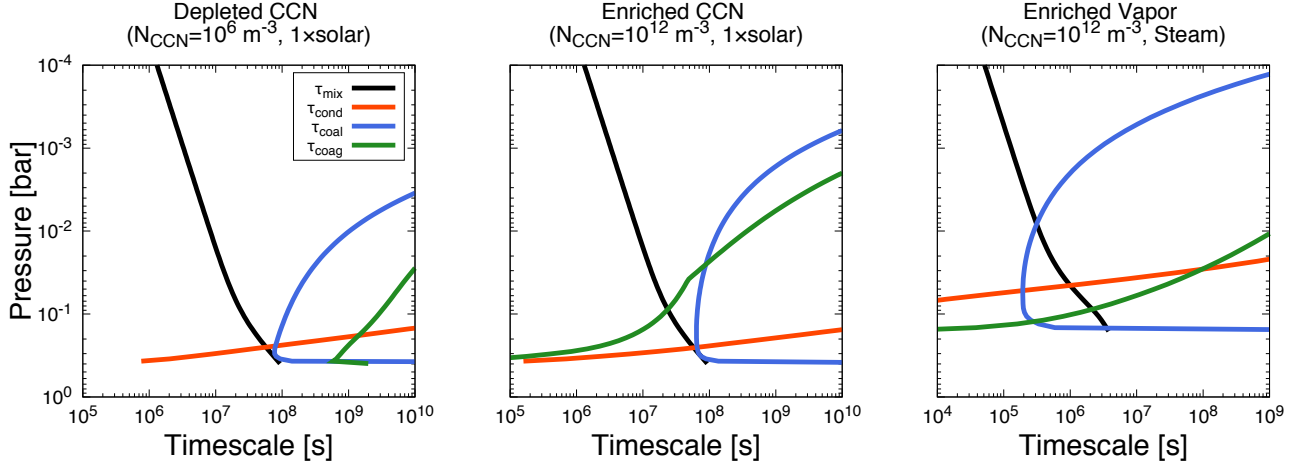}
\caption{Vertical distributions of the timescales of particle growth and vertical mixing. The left, middle, and right panels show the distributions for $n_{\rm CCN}={10}^{6}$ and ${10}^{12}~{\rm m^{-3}}$ with the metallicities of $1\times$ solar, and $n_{\rm CCN}={10}^{12}~{\rm m}^{-3}$ with the pure steam atmosphere. The black, red, blue, and green lines show the timescales of vertical mixing, coalescence, coagulation, and condensation respectively.}
\label{fig:timescale}
\end{figure*}
\begin{figure*}[t]
\includegraphics[clip,width=\hsize]{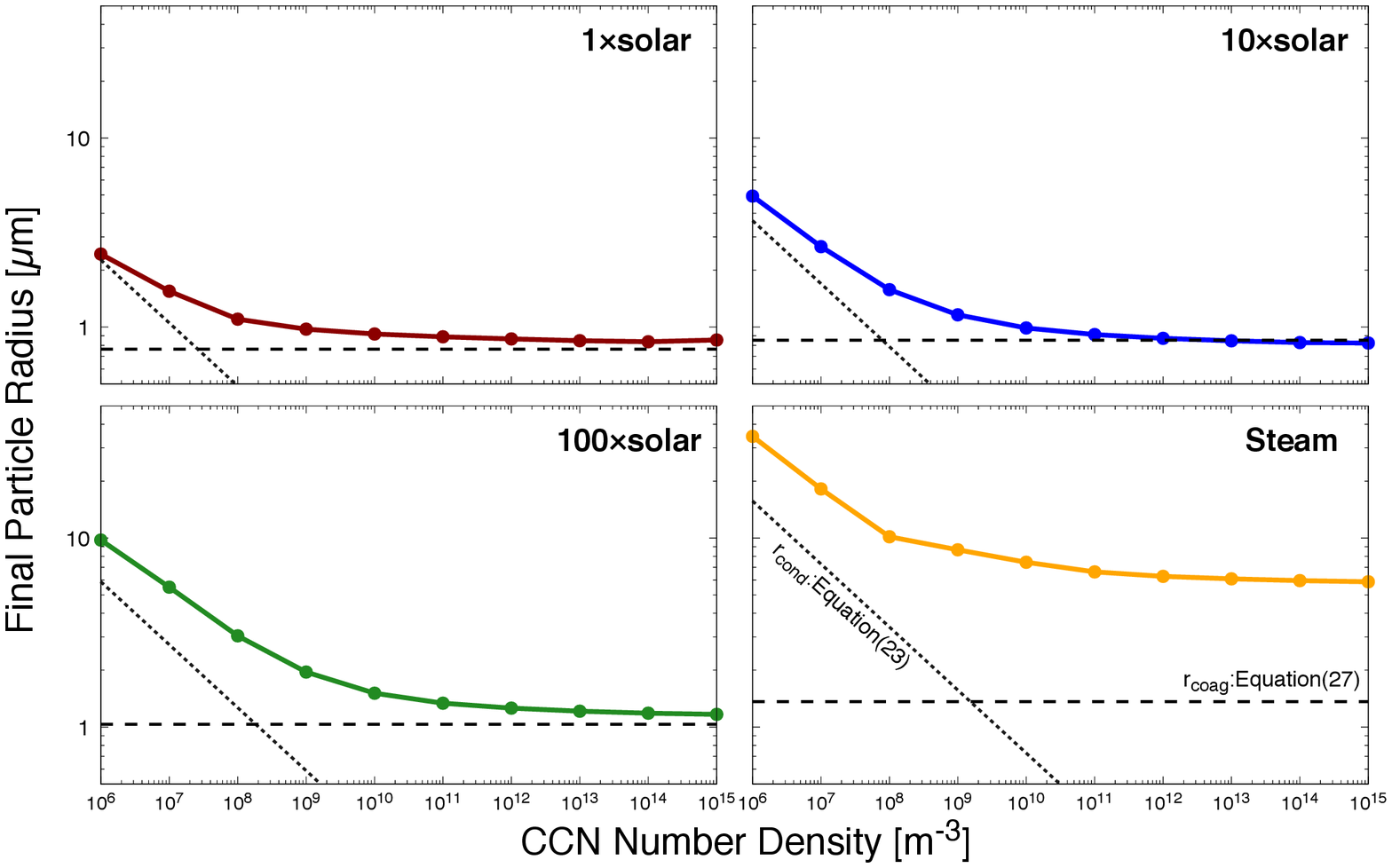}
\caption{Final particle radius as a function of the CCN number density. From top to bottom, each row shows the final radius for the metallicity of $1\times$, $10\times$, $100\times$ solar, and the steam atmosphere, respectively. The dashed and dotted lines show the size determined by coagulation, $r_{\rm coag}$, predicted by Equation \eqref{eq:r_coag2} and that by condensation, $r_{\rm cond}$, predicted by Equation \eqref{eq:r_cond}, respectively (see Section \ref{sec:cond-regime} and \ref{sec:coag-regime}).}
\label{fig:size_fin}
\end{figure*}
In this section, we particularly focus on the physical mechanisms that control the vertical distributions of the cloud particle size.
Figure \ref{fig:size} shows the calculated vertical profiles of mineral clouds in GJ1214 b.
We find that the particles grow mainly near the cloud base (left column in Figure \ref{fig:size}) and stop growing in the upper atmosphere where $P\leq{10}^{-3}~{\rm bar}$.
This occurs because the mixing timescale $\tau_{\rm mix}\propto K_{\rm z}^{-1} \propto P^{2/5}$ decreases with height, and eventually becomes shorter than the timescales of condensation, coagulation, and coalescence. 
This trends is also seen in the results of a recent cloud model that takes into account the evolution of particle size distribution \citep{Gao+18}.
The final particle radius ranges from $1$ to $2~{\rm \mu m}$ for the metallicity of $1\times$ solar, $0.9$ to $4~{\rm \mu m}$ for $10\times$ solar, $1.5$ to $10~{\rm \mu m}$ for $100\times$ solar, and $5$ to $30~{\rm \mu m}$ for water vapor atmosphere, respectively.
Figure \ref{fig:size} indicates that the final particle size decreases with the $n_{\rm CCN}$ and approaches a minimum value in the limit of high $n_{\rm CCN}$.
In Section \ref{sec:size_regime} we explain how the final particle size is determined.
We also find that a higher metallicity leads to a larger final size, although its effect is small compared to that of CCN number density.

The cloud mass mixing ratio, defined as $\rho_{\rm c}/\rho_{\rm g}$, steeply decreases with height above the height where $\tau_{\rm fall}<\tau_{\rm mix}$.
This can be understood from the transport equations.
In the upper atmosphere, the source terms expressing the particle growth are negligible as mentioned above.
Therefore, in a steady state, the vertical mixing of particles should balances with sedimentation,
 \begin{equation}\label{eq:balance}
-\rho_{\rm g}K_{\rm z}\frac{\partial }{\partial z}\left( \frac{\rho_{\rm c}}{\rho_{\rm g}}\right)-v_{\rm t}\rho_{\rm c}=0.
\end{equation}
When the $\tau_{\rm mix}\ll\tau_{\rm fall}$, Equation \eqref{eq:balance} indicates that $\rho_{\rm c}/\rho_{\rm g}$ is nearly constant for height, which is seen in the lower region of Figure \ref{fig:size}.
When the $\tau_{\rm mix}\gg\tau_{\rm fall}$, the mass mixing ratio decreases with height due to the particle sedimentation.

The vertical distribution of the cloud mass density also depends on the CCN number density and atmospheric metallicity (the right column of Figure \ref{fig:size}).
A larger CCN number density leads to a higher mass density at high altitude because the final particle size decreases with $n_{\rm CCN}$ as mentioned before.
We also find that a higher metallicity yields a higher cloud mass density at high altitude.
This metallicity dependence arises because the final particle size is insensitive to the metallicity, while the cloud mass density at the cloud base is approximately proportional to the metallicity.
Furthermore the dependence of $\tau_{\rm mix}\propto H^{2}$ also yields the higher cloud mass at high altitude for higher metallicity cases because the $H$ decreases with increasing atmospheric metallicity.

\subsection{The Mechanisms Controlling Particle Size}\label{sec:size_regime}
The final particle size determines how high the cloud particles can ascend.
Here we discuss the mechanisms that control the final particle size. 
Figure \ref{fig:timescale} shows the vertical distributions of the timescales of vertical mixing, condensation, coagulation, and coalescence for three cases: depleted CCNs ($n_{\rm CCN}={10}^{6}~{\rm m^{-3}}$), enriched CCNs ($n_{\rm CCN}={10}^{12}~{\rm m^{-3}}$), and enriched vapor (steam atmospheres).
The timescales of condensation, coagulation, and coalescence are defined as
\begin{equation}
\tau_{\rm cond}=\rho_{\rm c}\left| \frac{\partial \rho_{\rm c}}{\partial t}\right|_{\rm cond}^{-1},
\end{equation}
\begin{equation}\label{eq:t_coag_general}
\tau_{\rm coag}=n_{\rm c}\left| \frac{\partial n_{\rm c}}{\partial t}\right|_{\rm coag}^{-1},
\end{equation}
\begin{equation}\label{eq:t_coal_general}
\tau_{\rm coal}=n_{\rm c}\left| \frac{\partial n_{\rm c}}{\partial t}\right|_{\rm coal}^{-1},
\end{equation}
and the mixing timescale $\tau_{\rm mix}$ is given by Equation \eqref{eq:t_mix}.
Generally, cloud particles grow if ${\rm min}(\tau_{\rm cond},\tau_{\rm coag},\tau_{\rm coal})\ll \tau_{\rm mix}$, and ascend without significant growth if ${\rm min}(\tau_{\rm cond},\tau_{\rm coag},\tau_{\rm coal})\gg \tau_{\rm mix}$.
The mixing timescale $\tau_{\rm mix}$ decreases with height as mentioned before, whereas the growth timescales increase with height because they are inversely proportional to the density.
Hence the particle growth becomes relatively less effective as the particles ascend.
In following subsections, we characterize the particle growth in three cases based on timescale argument.

\subsubsection{Depleted CCN Regime ($\tau_{\rm mix}<\tau_{\rm coag}, \tau_{\rm coal}$)}\label{sec:cond-regime}
In the example shown in the left panel of Figure \ref{fig:timescale}, $\tau_{\rm cond}$ is much shorter than $\tau_{\rm mix}$ and other growth timescales at the cloud base.
The short $\tau_{\rm cond}$ results in the quick growth of particles near the cloud base as shown in Figure \ref{fig:size}.
At the same time the rapid condensation also results in rapid depletion of condensing vapor. 
This depletion eventually suppresses the condensation growth,
and hence the total cloud mass at the cloud base is limited by the total amount of condensing vapor there, i.e., $\rho_{\rm c}(z_{\rm b})\approx \rho_{\rm v}(z_{\rm b})=\rho_{\rm s}(z_{\rm b})$.

If $n_{\rm CCN}$ is so small that ${\rm min}( \tau_{\rm coal}, \tau_{\rm coag})>\tau_{\rm mix}$ at the cloud base, the particles start to ascend as soon as the condensation growth is completed (the left column in Figure \ref{fig:timescale}).
In this case, the final particle size $r_{\rm cond}$ is determined by the deposition of available vapor onto CCNs, i.e.,
\begin{equation}
\frac{4}{3}\pi r_{\rm cond}^{3}\rho_{\rm p}n_{\rm CCN}\approx \rho_{\rm s}(z_{\rm b}),
\end{equation}
and thus
\begin{equation}\label{eq:r_cond}
r_{\rm cond}\approx \left[ \frac{3\rho_{\rm s}(z_{\rm b})}{4\pi \rho_{\rm p}n_{\rm CCN}}\right]^{1/3},
\end{equation}
where we have assumed the initial CCN mass density is much smaller than $\rho_{\rm s}(z_{\rm b})$.
Figure \ref{fig:size_fin} shows the final particle size and $r_{\rm cond}$ for each metallicity case.
As shown in Figure \ref{fig:size_fin}, the final particle size approaches $r_{\rm cond}$ for lower CCN number density.
Hence Equation \eqref{eq:r_cond} explains why the final particle size decreases with the increasing of CCN number density.
Equation \eqref{eq:r_cond} also explains the results of \citet{Gao+18}, who found that the efficient homogeneous nucleation (high particle number density) results in small particle size.

\subsubsection{Enriched CCN Regime ($\tau_{\rm coag}<\tau_{\rm mix}<\tau_{\rm coal}$)}\label{sec:coag-regime}
Coagulation leads the further growth of cloud particles in addition to condensation if $n_{\rm CCN}$ is so high that $\tau_{\rm coag}<\tau_{\rm mix}$ at the cloud base (see middle panel of Figure \ref{fig:timescale}).
When coagulation is effective, the final particle size becomes larger than $r_{\rm cond}$ and eventually reaches the minimum value in the limit of high CCN number density as seen in Figure \ref{fig:size}.
Hence we can expect the particle size must be larger than the minimum value determined by coagulation even if the CCN number density is uncertain.

The minimum particle size can be analytically estimated in the following way.
Because the final particle size ranges as $r>0.07~{\rm \mu m}$ in Figure \ref{fig:size}, the coagulation growth falls into diffusive regime, and the $\tau_{\rm coag}$ is written by
\begin{equation}\label{eq:t_coag}
\tau_{\rm coag}=\frac{3\eta}{4k_{\rm B}T\beta n_{\rm c}}.
\end{equation}
Also the slip factor can be approximated as $\beta \approx \beta_{\rm \infty}{\rm Kn}_{\rm g}$, where $\beta_{\rm \infty}=1.657$, 
because the mean free path near the cloud base ($l\sim10~{\rm \mu m}$) is larger than the particle radius, i.e., ${\rm Kn_g}\gg1$.
Using the relation $4\pi r_{\rm c}^3\rho_{\rm p}n_{\rm c}/3=\rho_{\rm c}$ and $\eta=\rho_{\rm g}v_{\rm th}l/3$ \citep{Woitke&Helling03}, where $v_{\rm th}=\sqrt{8k_{\rm B}T/\pi m_{\rm g}}$ is the mean thermal velocity, the coagulation timescale can be rewritten as
\begin{eqnarray}\label{eq:t_coag0}
\nonumber
\tau_{\rm coag}&=&\frac{\rho_{\rm g}v_{\rm th}r_{\rm c}}{4k_{\rm B}T\beta_{\rm \infty} n_{\rm c}}\\
&=&\frac{\pi\rho_{\rm p}v_{\rm th}}{3k_{\rm B}T\beta_{\rm \infty} q_{\rm c}}r_{\rm c}^4,
\end{eqnarray}
where $q_{\rm c}\equiv \rho_{\rm c}/\rho_{\rm g}$ is the cloud mass mixing ratio.
Because $\rho_{\rm c}$ is determined by saturation vapor density at the cloud base (see Section \ref{sec:cond-regime}), $q_{\rm c}=\rho_{\rm s}(z_{\rm b})/\rho_{\rm g}(z_{\rm b})=m_{\rm KCl}q_{\rm KCl}/m_{\rm g}$, where $m_{\rm KCl}$ is the mass of a KCl molecule.
Coagulation growth completes when the vertical mixing becomes more efficient, and hence the final size is determined from the condition $\tau_{\rm coag}=\tau_{\rm mix}$.
Equating Equations \eqref{eq:t_coag0} and \eqref{eq:t_mix}, the final particle size determined by coagulation $r_{\rm coag}$ is predicted as
\begin{equation}\label{eq:r_coag}
r_{\rm coag}=\left( \frac{3\beta_{\rm \infty}}{\sqrt{8\pi}}\frac{m_{\rm KCl}q_{\rm KCl}}{\rho_{\rm p}K_{\rm z}(z_{\rm b})}g^{1/2}H^{5/2} \right)^{1/4}\left(\frac{P_{\rm *}}{P_{\rm b}}\right)^{1/10},
\end{equation}
where $P_{\rm b}$ is the pressure of the cloud base and $P_{\rm *}$ is the pressure in which the coagulation growth is completed.
Equation \eqref{eq:r_coag} implies the final particle size in this regime is almost independent of $n_{\rm CCN}$ because $P_{\rm *}$ is insensitive to the choice of $n_{\rm CCN}$ as seen in Figure \ref{fig:timescale}.
This explains why the final particle size is almost independent of $n_{\rm CCN}$ for high CCN number density in Figure \ref{fig:size}. 
Particularly, we find that Equation \eqref{eq:r_coag} is in a good agreement with the minimum final size derived from the numerical results if we assume $P_{\rm *}=0.1P_{\rm b}$.
In this case, Equation \eqref{eq:r_coag} can be rewritten as the following useful formula
\begin{eqnarray}\label{eq:r_coag2}
r_{\rm coag}&=&1.25~{\rm \mu m}\times \\
\nonumber
&& \left( \frac{g}{10~{\rm m~s^{-2}}}\right)^{1/8}\left( \frac{H}{{10}^2~{\rm km}}\right)^{5/8}\left( \frac{K_{\rm z}(z_{\rm b})}{{10}^3~{\rm m^2~s^{-1}}}\right)^{-1/4}\left( \frac{q_{\rm KCl}}{{10}^{-5}}\right)^{1/4}.
\end{eqnarray}
Figure \ref{fig:size_fin} shows that the final particle size asymptotically reaches that predicted from Equation \eqref{eq:r_coag2} except the case of steam atmosphere.
The deviation for steam atmosphere is caused by coalescence as explained in next subsection.

\subsubsection{Enriched Vapor Regime ($\tau_{\rm coal}< \tau_{\rm mix}$)}\label{sec:coal-regime}
Coalescence is dominant only if condensing vapor is very abundant as in pure steam atmospheres as shown in the right panel of Figure \ref{fig:timescale}.
When coalescence is dominant, the final particle size becomes larger than the lower limit set by coagulation $r_{\rm coag}$ (see the bottom panel of Figure \ref{fig:size_fin}).
Because larger particles have larger settling velocity, coalescence suppresses the cloud-top height in the steam atmosphere. 
 
Here we predict the threshold abundance of condensing vapor that induces the significant growth through coalescence.
Because the particle size is larger than the gas mean free path near the cloud base $\sim 10~{\rm \mu m}$ in most of our calculations, 
the terminal velocity is expressed as the Epstein's law, approximated as
\begin{equation}\label{eq:vt_eps}
v_{\rm t}(r_{\rm c})\approx\frac{2\beta_{\rm \infty}g\rho_{\rm p}}{3\rho_{\rm g}v_{\rm th}}r_{\rm c}.
\end{equation}
Therefore the coalescence timescale can be rewritten as
\begin{eqnarray}\label{eq:t_coal}
\nonumber
\tau_{\rm coal}&=&\frac{1}{2\pi r^2 \Delta v n}\\
&=&\frac{2 v_{\rm th}}{\beta_{\rm \infty} gq_{\rm c}}.
\end{eqnarray}
Because $q_{\rm c}=m_{\rm KCl}q_{\rm KCl}/m_{\rm g}$ (see Section \ref{sec:coag-regime}), the coalescence timescale just above the cloud base is independent of $n_{\rm CCN}$, and only depends on the mixing ratio of the condensing vapor.
If $\tau_{\rm coal}\ll \tau_{\rm mix}$, the cloud particles grows via coalescence in addition to condensation and coagulation.
Comparing Equation \eqref{eq:t_coal} with $\tau_{\rm mix}(z_{\rm b})$, we find that coalescence occurs near the cloud base if the condensate mixing ratio is much higher than
\begin{eqnarray}\label{eq:q_cri}
q_{\rm *}&\approx& \frac{2v_{\rm th}K_{\rm z}m_{\rm g}}{\beta_{\rm \infty}gH^2m_{\rm KCl}}
\nonumber \\
&\approx& 5\times{10}^{-7} \left( \frac{g}{10~{\rm m~s^{-2}}}\right)^{-1/2}\left( \frac{H}{{10}^2~{\rm km}}\right)^{-3/2}
\nonumber \\
&&\times \left( \frac{K_{\rm z}(z_{\rm b})}{10^{3}~{\rm m^2~s^{-1}}}\right)\left( \frac{m_{\rm g}}{2~{\rm amu}}\right).
\end{eqnarray}

Substituting the parameters for the steam atmosphere of GJ1214 b, the mixing ratio of condensing vapor $q_{\rm KCl}=2.61\times{10}^{-4}$ exceeds the $q_{\rm *}\sim 2\times{10}^{-5}$ by an order of magnitude, and hence coalescence dominates the particle growth.

\subsection{Predicting the Maximum Cloud-Top Height}\label{sec:cloudtop}
\begin{figure}
\includegraphics[clip,width=\hsize]{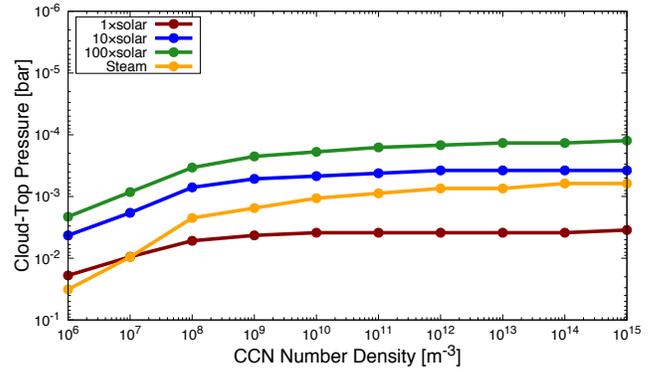}
\caption{Cloud-top pressure as a function of CCN number density for different atmosphere models. The red, blue, green, and yellow lines are for $1\times$, $10\times$, $100\times$ solar models, and pure steam atmosphere, respectively. 
}
\label{fig:cloudtop}
\end{figure}

The prediction of the cloud-top height, defined as the height at which the atmosphere becomes opaque due to the cloud, is important because it determines the shape of observed spectra \citep[e.g.,][]{Brown01}.
In order to predict it, we calculate the slant optical depth $\tau_{\rm s}$, defined as the optical depth for the path length of the transmitted starlight, using Equation (6) of \citet{Fortney05}, given by
\begin{equation}\label{eq:Fortney}
\tau_{\rm s}=\tau_{\rm v}\sqrt{\frac{2\pi R_{\rm p}}{H_{\rm c}}},
\end{equation}
where $\tau_{\rm v}$ is the vertical optical depth of the cloud, $R_{\rm p}$ is the planetary radius, and 
$H_{\rm c} = |d{\rm ln}n_{\rm c}/dz|^{-1}$ is the cloud scale height.
The cloud-top height can be estimated as the height at which $\tau_{\rm s}$ exceeds unity.
The vertical optical depth $\tau_{\rm v}$ is given by
\begin{equation}\label{eq:tau}
\tau_{\rm v}(z)=\int^{\infty}_{z}Q_{\rm ext}(r_{\rm c})\pi r_{\rm c}^2n_{\rm c}dz,
\end{equation}
where $Q_{\rm ext}$ is the extinction coefficient of the particles.
To calculate $Q_{\rm ext}$, we perform rigorous Mie calculations using {\it BHMIE} code \citep{Bohren&Huffman83}.
We use the refractive index of KCl from \citet{Querry87} and assume an isotropic scattering for the calculations of scattering opacities.
We assume the wavelength of $\lambda=1.4~{\rm \mu m}$, at which a prominent water feature is located.
For GJ1214 b and GJ436 b, this feature is absent in the actual spectra \citep{Kreidberg+14a,Knutson+14a}, and therefore the cloud-top height defined at $1.4~{\rm \mu m}$ must be sufficiently high so that the clouds fully obscure the feature.

To estimate $H_{\rm c}$, we use the fact that at high altitudes particle growth is negligible and hence the vertical profiles are given by Equation \eqref{eq:balance}.
Since $H = |d{\rm ln}n_{\rm g}/dz|^{-1}$ and $H_{\rm c} = |d{\rm ln}n_{\rm c}/dz|^{-1}$, 
Equation \eqref{eq:balance} can be rewritten as
\begin{equation}\label{eq:Hc}
H_{\rm c}=H\left(1+\frac{v_{\rm t}H}{K_{\rm z}}\right)^{-1}.
\end{equation}
Because the term $v_{\rm t}H/K_{\rm z}$ is the ratio of $\tau_{\rm mix}$ to $\tau_{\rm fall}$, $H_{\rm c} \approx  H$ for $\tau_{\rm mix}\ll \tau_{\rm fall}$
and $K_{\rm c} \approx K_{\rm z}/v_{\rm t}$ for $\tau_{\rm mix}\gg \tau_{\rm fall}$.

Figure \ref{fig:cloudtop} shows the cloud-top pressure for GJ1214 b predicted by our calculations for different values of the metallicity and $n_{\rm CCN}$.
We find that the cloud-top height increases with $n_{\rm CCN}$, but plateaus in the limit of high $n_{\rm CCN}$.
This means that one can predict the maximum height of the cloud top for given atmospheric metallicity.
The presence of the maximum height results from the presence of the minimum particle size mentioned in Section \ref{sec:coag-regime}.
The impact of size distribution on the predicted maximum height is small as we will discuss in Section \ref{sec:sizedist}.

We also find that metal-enriched atmospheres are more likely to yield vertically extended clouds for the abundance of condensing vapor below the threshold (Equation \eqref{eq:q_cri}).
Figure \ref{fig:cloudtop} shows that the cloud-top is placed at $P\ga3\times{10}^{-3}~{\rm bar}$ for $1\times$ solar metallicity, $P\ga3\times{10}^{-4}~{\rm bar}$ for $10\times$ solar metallicity, and $P\ga 1\times{10}^{-4}~{\rm bar}$ for $100\times$ solar metallicity, and $P\ga 6\times{10}^{-4}~{\rm bar}$ for the steam atmosphere.
This trend arises because a higher metallicity atmosphere yields a higher total cloud mass and more efficient vertical mixing as mentioned before.
For vapor abundance above the threshold, the case of the steam atmosphere, cloud-top height no longer increases with metallicities because coalescence leads significant growth for cloud particles as mentioned in Section \ref{sec:coal-regime}.

\section{Application to GJ1214 b and GJ436 b}\label{sec:observation}
\begin{figure*}
\includegraphics[clip,width=\hsize]{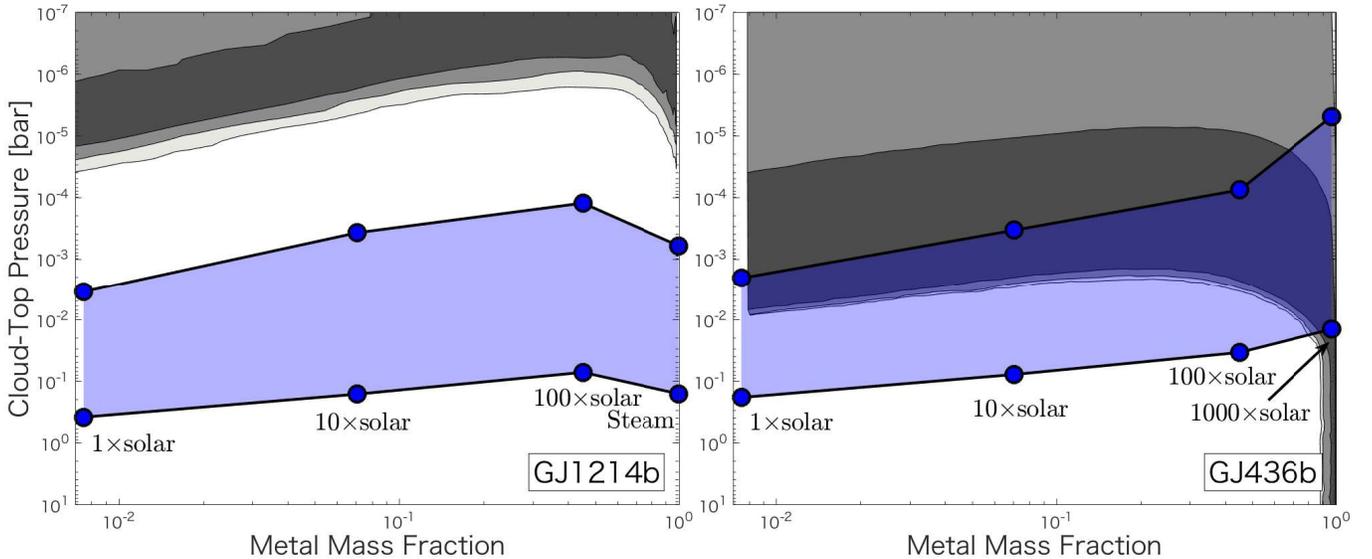}
\caption{Predicted maximum extent of the KCl cloud for GJ1214 b (left panel) and GJ436 b (right panel) as a function of the metal mass fraction.
The dots correspond to, from left to right, hydrogen-rich atmosphere at $1\times$, $10\times$, $100\times$ solar metallicity, and pure steam atmosphere for GJ1214 b and $100\times$ solar metallicity for GJ436 b, respectively.
The lower line indicates the height (in pressure) of the cloud base, while the upper line indicates the maximum height of the cloud top for fixed metallicity.
The gray shaded area indicates the location of the cloud top inferred from the Bayesian analysis of the transmission spectrum by \citet{Kreidberg+14a} for GJ1214 b and by \citet{Knutson+14a} for GJ436 b, with the black contours marking the $1\sigma$, $2\sigma$, and $3\sigma$ Bayesian credible regions.
}
\label{fig:GJ1214b}
\end{figure*}

Now we apply our cloud model to two super-Earths, GJ1214 b and GJ436 b, which are known to exhibit a flat transmission spectrum.
As mentioned in the previous section, there is a maximum height, or equivalently a minimum atmospheric pressure $P_{\rm min}$, that can be reached by the top of a KCl cloud for given atmospheric metallicity.
In order to examine whether KCl clouds are responsible for the flat transmission spectra, 
we compare the maximum cloud-top heights with the cloud heights observationally inferred for the two super-Earth.

\subsection{GJ1214 b}
In the left panel of Figure \ref{fig:GJ1214b}, we plot the height (in pressure) of the cloud base and the maximum height of the cloud top predicted for GJ1214b as a function of the metal mass fraction of the atmosphere\footnote{Metal mass fraction is defined as the mass fraction of heavy element. Following \citet{Fortney+13}, we calculate metal mass fraction assuming H--He--water mixtures in this study.}.
The blue-shaded area in Figure \ref{fig:GJ1214b} thus indicates the heights where 
the top of the KCl cloud can exist  for some CCN number density. For comparison, we also indicate by the gray-shaded the heights of the cloud top suggested by \citet{Kreidberg+14a} 
based on Bayesian analysis on the observed transmission spectrum.

We find that the maximum cloud-top height is too low to explain the flat spectrum 
for all plausible values of the atmospheric metallicity.
In principle, a higher atmospheric metallicity provides a higher cloud-top height as already mentioned in Section \ref{sec:cloudtop}.  
However, even if we assume the steam atmosphere, the maximum cloud-top height ($P_{\rm min}=6\times{10}^{-4}~{\rm bar}$ in pressure) is still an order of magnitude higher than inferred by \citet{Kreidberg+14a} (cloud-top pressure $\leq 3\times {10}^{-5}$ bar at $3\sigma$ confidence).
This is because in the steam atmosphere, coalescence causes the significant growth of cloud particles of $r_{\rm c}\geq5~{\rm \mu m}$.

The above comparison clearly shows that a simple condensate cloud cannot explain the flat transmission spectrum of GJ1214 b.
This fact might support the idea that the flat spectrum of GJ1214 b is caused by photochemical haze \citep{Miller-Ricci+12,Morley+13,Morley+15,Kawashima&Ikoma18} rather than by mineral clouds.
Alternatively, our cloud model might still be missing important physics of particle growth. 
For example, it is suggested both theoretically and experimentally \citep[e.g.,][]{Dominik&Tielens97,Blum&Wurm00} that, unlike water cloud droplets, solid particles grow into 
highly porous particles through mutual sticking. 
This porosity evolution is neglected in Figure \ref{fig:GJ1214b}, but could help particles ascend 
to very high altitudes because porous particles have a 
lower settling velocity than compact particles of the same mass. 
We address this possibility in Section \ref{sec:porosity}.

\subsection{GJ436 b}
For GJ436 b, we find that the maximum cloud-top height for KCl clouds is high enough to be 
consistent with the transmission observations.
The Bayesian analysis by \citet{Knutson+14a} indicates that the cloud top is present 
 at atmospheric pressures of $\la 10^{-2}~\rm bar$ except for metal-rich atmospheres of metal mass fraction $\ga 0.8$ for which the location of the cloud top is not well constrained (see the gray-shaded area in the right panel of Figure~\ref{fig:GJ1214b}). 
As shown in the right panel of Figure \ref{fig:GJ1214b}, 
the minimum cloud-top pressure $P_{\rm min}$ predicted from our model
is $2\times{10}^{-3}$, $3\times{10}^{-4}$, $8\times10^{-5}$, and $5\times{10}^{-6}~{\rm bar}$ for the metallicities of $1\times$, $10\times$, $100\times$, and $1000\times$ solar, respectively. 
Since we adopted the high $K_{\rm z}$ for the metallicity of $1000\times$ solar compared to that for the steam atmosphere on GJ1214 b, cloud particles avoid the significant growth due to coalescence.
This is a reason why the cloud-top height for $1000\times$ solar is much higher than that for the steam atmospheres on GJ1214 b.
Combining the Bayesian analysis results and our model prediction, we suggest that the flat spectrum of GJ436 b is likely caused by a KCl cloud with its top at $ \sim 10^{-3}$--$10^{-2}$ bar for hydrogen-rich atmospheres (metal mass fraction $\la 0.8$) and at $\sim 10^{-2}$--$10^{-5}$ bar for metal-rich atmospheres (metal mass fraction $\ga 0.8$). 
However, because we here adopted a metallicity-independent eddy diffusion coefficient 
(see Section \ref{sec:transport}), we cannot conclude whether the atmosphere of GJ436 b is likely 
to be hydrogen-rich or metal-rich. Future three-dimensional modeling of GJ436 b's atmospheric circulation, like the one done by \citet{Charnay+15} for GJ1214 b, would allow us to determine the atmosphere's metallicity.

\section{Discussion} \label{sec:discussion}
\subsection{Influences of Size Distribution on the Cloud-Top Height}\label{sec:sizedist}
\begin{figure}
\centering
\includegraphics[clip,width=0.95\hsize]{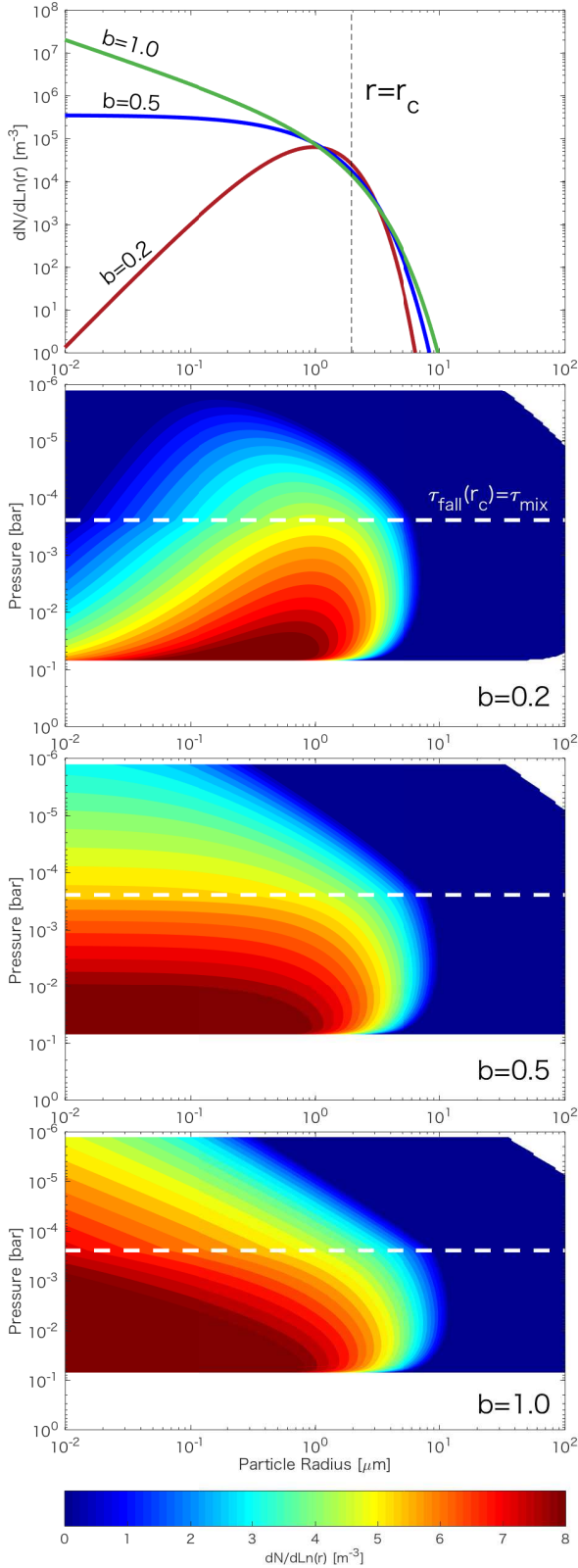}
\caption{Constructed particle size distributions. The metallicity of $100\times$ solar and $N_{\rm CCN}={10}^{9}~{\rm m^{-3}}$ are selected. Top panel shows the size distributions for $b=0.1$, $0.5$, and $1.0$ at the height where $\tau_{\rm mix}=\tau_{\rm fall}(r_{\rm c})$, denoted as white dotted lines in lower panels.
Each panel, from second to bottom, shows the vertical size distributions for $b=0.1$, $0.5$, and $1.0$, respectively. 
}
\label{fig:sizedis}
\end{figure}

\begin{figure*}[t]
\includegraphics[clip,width=\hsize]{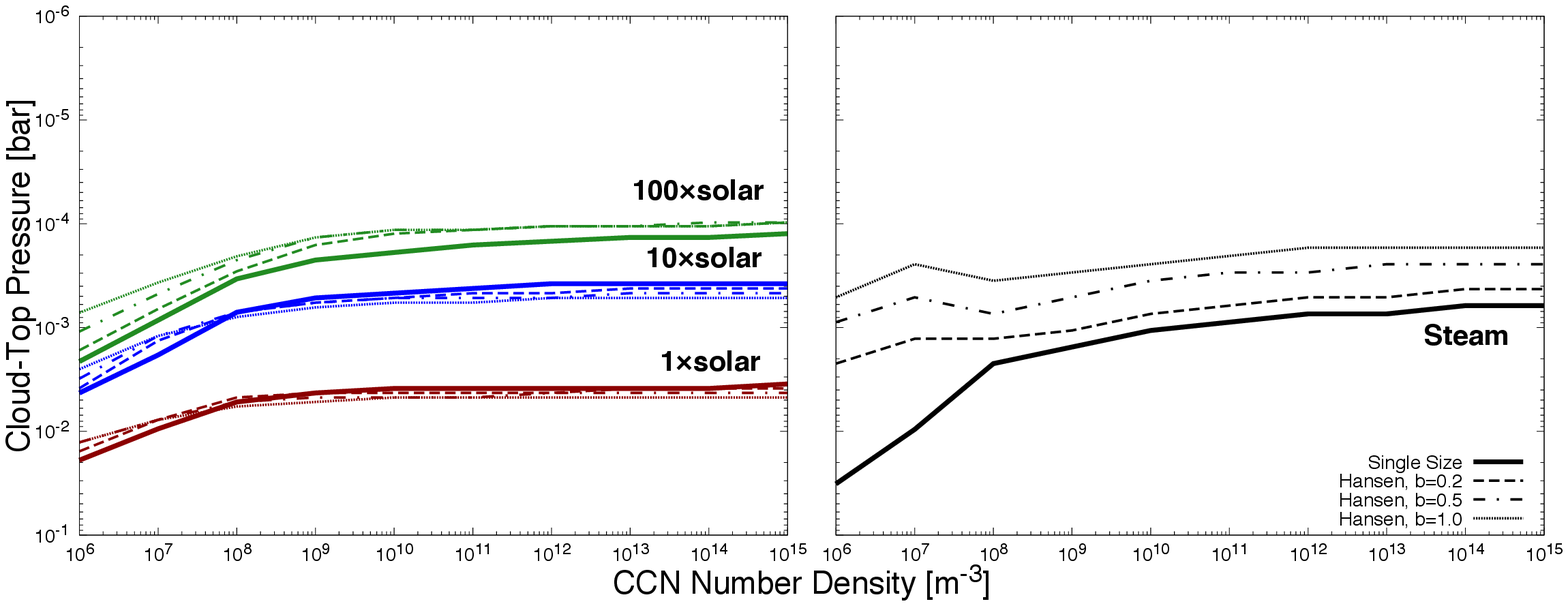}
\caption{Cloud-top pressures as a function of CCN number density for hydrogen-rich atmosphere models (left panel) and the steam atmosphere model (right panel) obtained from models of different particle size distribution.
The red, blue, green, and black lines are for $1\times$, $10\times$, $100\times$ solar models, and pure steam atmosphere, respectively. The solid lines are calculated for characteristic size method, while the dashed lines, dashed dotted lines, and dotted lines are calculated for the Hansen size distribution with $b=0.2$, $0.5$, and $1.0$, respectively.
}
\label{fig:cloudtop_size}
\end{figure*}

Because the total particle cross section tends to be dominated by small particles rather than by the particles dominating the total cloud mass, the cloud-top height might be influenced by the size distribution, which is however not captured by our calculations. 
We here evaluate the impact of particle size distribution on the predicted cloud-top height by adding to our model a distribution of small particles.
We assume that particles smaller than $r_{\rm c}$ obey the Hansen size distribution \citep{Hansen71} given by
\begin{equation}\label{eq:hansen}
f(r)\equiv \frac{dn}{dr}=Cr^{(1-3b)/b}\exp{\left(-\frac{r}{ab}\right)},
\end{equation}
where $f(r)dr$ is the number density of particles with radii between $r$ and $r+dr$, $C$ is a constant, $a$ and $b$ are the mean effective radius and the effective variance defined by
\begin{equation}
a\equiv\frac{\int^{\infty}_{0} r\pi r^2 f(r)dr}{\int^{\infty}_{0} \pi r^2f(r)dr},
\end{equation}
\begin{equation}
b\equiv\frac{\int^{\infty}_{0} (r-a)^2\pi r^2 f(r)dr}{a^2\int^{\infty}_{0} \pi r^2f(r)dr}.
\end{equation}
The Hansen size distribution successfully reproduces the observed size distributions of terrestrial water clouds for $b=0.1$--$0.2$ \citep{Hansen71}, and near-infrared spectral energy distributions of cloudy brown dwarfs for $b>0.5$ \citep{Hiranaka+16}.
The top panel of Figure \ref{fig:sizedis} shows the Hansen size distributions for $b=0.2$, $0.5$, and $1.0$.
One can see that $b<0.5$ yields log-normal like size distributions, while $b>0.5$ yields power-law like size distributions.
Therefore, the Hansen size distribution with various choices of $b$ enables us to test the size distributions of a various shape.

For each height, we determine the $a$ and $C$ so that the mass weighted size and the cloud mass density correspond to $r_{\rm c}$ and $\rho_{\rm c}$ calculated by our model, respectively.
We calculate the $a$ and $C$ at each height using the following relations,
\begin{equation}
r_{\rm c}=\frac{\int ^{\infty}_{0}rm(r)f(r)dr}{\int ^{\infty}_{0}m(r)f(r)dr}=a(1+b),
\end{equation}
\begin{equation}
\rho_{\rm c}=\frac{4\pi \rho_{\rm p}}{3}C(ab)^{(1+b)/b}\Gamma \left(\frac{1+b}{b}\right),
\end{equation}
where $\Gamma(z)$ is the gamma function.
However, $r_{\rm c}$ might be overestimated in our calculations at the height where $\tau_{\rm mix}<\tau_{\rm fall}(r_{\rm c})$ because our model fails to trace the decreasing of mass weighted size due to the removal of large particles by gravitational settling.
To avoid this issue, we use an analytical solution of the transport equation.
In the upper atmosphere, the particle growth is negligible as mentioned in Section \ref{sec:result}, and hence, in a steady state, the particle number density obeys 
\begin{equation}\label{eq:balance_n}
-n_{\rm g}K_{\rm z}\frac{\partial}{\partial z}\left( \frac{n_{\rm c}}{n_{\rm g}}\right)-v_{\rm t}n_{\rm c}=0.
\end{equation}
If we approximate $\beta=1+\beta_{\rm \infty}{\rm Kn_{\rm g}}$\footnote{This expression asymptotically approaches Equation \eqref{eq:slip} in the limits of small and large $\rm {Kn}_{\rm g}$. The maximum deviation from Equation \eqref{eq:slip} is only $\approx10\%$, which occurs at ${\rm Kn}_{\rm g}=1$.}, Equation \eqref{eq:balance_n} can be analytically solved as
\begin{eqnarray}\label{eq:balance0}
n_{\rm c}(P)&=&n_{\rm c}(P_{\rm 0})\mathcal{P}\times\\ 
\nonumber
&&\exp{\left[ \frac{5\chi(P_{\rm 0})}{2(1+\beta_{\infty}{\rm Kn}_{\rm 0})}\left( ( \mathcal{P}^{2/5}-1) -\frac{2 \beta_{\rm \infty}{\rm Kn_{\rm 0}}}{3}(\mathcal{P}^{-3/5}-1) \right)\right]},
\end{eqnarray}
where $\mathcal{P}\equiv P/P_{\rm 0}$ and $\chi(P)\equiv \tau_{\rm mix}(P)/\tau_{\rm fall}(P)$ is the ratio of the mixing timescale to the falling timescale.
We calculate the size distributions at the regions of $\tau_{\rm mix}<\tau_{\rm fall}(r_{\rm c})$ using Equation \eqref{eq:balance0} for each size bin.
Figure \ref{fig:sizedis} shows the constructed vertical size distributions for $b=0.2$, $0.5$, and $1.0$.
The reference pressure $P_{\rm 0}$ is set as a height where $\tau_{\rm mix}=\tau_{\rm fall}(r_{\rm c})$, denoted as the white dotted lines in each panel.
Figure \ref{fig:sizedis} indicates that the larger $b$ is, the more small particles are present at high altitude.

Figure \ref{fig:cloudtop_size} compares the cloud-top heights predicted by the characteristic size model with those by the model with the Hansen size distribution of $b=0.2$, $0.5$, and $1.0$.
With particle size distribution, the vertical optical depth is calculated as
\begin{equation}
\tau_{\rm v}(z)=\int_{z}^{\infty}\int_{0}^{\infty}Q_{\rm ext}(r)\pi r^2 f(r,z)drdz.
\end{equation}
We find that size distribution has little effect on the cloud top height except for the case of the steam atmosphere.
This is because KCl is a purely scattering material in near-infrared, i.e., the extinction is equivalent to the scattering.
For purely scattering particles smaller than the wavelength, the extinction efficiencies steeply decreases with decreasing the particle size as $Q_{\rm ext}\propto r^{4}$.
Therefore, the contribution of such small particles ($r\ll \lambda$) to the total cloud opacity is negligibly small.
The most efficient extinction occurs at $r\sim \lambda/2\pi$, which is $0.2~{\rm \mu m}$ for $\lambda=1.4~{\rm \mu m}$.
By contrast, the final characteristic sizes for the metallicities of $1\times$, $10\times$, and $100\times$ solar are $r_{\rm c}\approx1~{\rm \mu m}$ (see Figure \ref{fig:size_fin}), already close to $0.2~{\rm \mu m}$. 
Therefore, the addition of particles smaller than $r_{\rm c}$ has little effect on the optical depth, and hence on the cloud-top height.

The difference arising from size distributions becomes obvious only when the final characteristic size is orders of magnitude larger than $r=\lambda/2\pi$.
This is the case for the steam atmospheres, in which the final characteristic size is $r\approx5~{\rm \mu m}\gg0.2~{\rm \mu m}$.
In this case, varying the size distribution can decrease the cloud-top pressure by a factor of $3$ from the prediction of the characteristic size method.
However, we find that the cloud-top height for GJ1214 b with a steam atmosphere is still an order of magnitude lower than anticipated from the observation of \citet{Kreidberg+14a}.
Therefore, we conclude that one cannot explain the flat spectrum of GJ1214 b solely by considering particle size distribution.

\subsection{Influences of the Convective Adjustment}\label{sec:PT_dis}
\begin{table}[t]
  \centering
   \caption{Maximum Cloud-Top Height for Different P-T structure}
  \begin{tabular}{c|cc}
     metallicity & \citet{Charnay+15} & \citet{Guillot10} \\ \hline
     1$\times$solar &  $P_{\rm min}=3.6\times{10}^{-3}~{\rm bar}$ & $P_{\rm min}=3.5\times{10}^{-3}~{\rm bar}$ \\ 
    10$\times$solar & $P_{\rm min}=4.7\times{10}^{-4}~{\rm bar}$ & $P_{\rm min}=3.8\times{10}^{-4}~{\rm bar}$  \\ 
        100$\times$solar & $P_{\rm min}=1.4\times{10}^{-4}~{\rm bar}$ & $P_{\rm min}=1.2\times{10}^{-5}~{\rm bar}$  \\
     Steam & $P_{\rm min}=1.2\times{10}^{-3}~{\rm bar}$  & $P_{\rm min}=6.2\times{10}^{-4}~{\rm bar} $
  \end{tabular}
  \label{table:3}
\end{table}

Our P-T structure neglects heat transport by convection, which is the process so called convective adjustment and in reality becomes important when the temperature steeply declines with decreasing pressure \citep{Manabe&Stricler64,Marley&Robinson15}.
We performed test calculations using the P-T structure provided by \citet{Charnay+15}, which includes the effect of the convective adjustment.
As listed in Table \ref{table:3}, we confirmed that the maximum heights of cloud top are quantitively similar to the results from radiative P-T profiles of \citet{Guillot10}.
The largest influence is only a factor of $2$, which occurs for the case of the steam atmospheres.
The reason why the cloud-top height is nearly same for the both P-T structures is that the convective adjustment only changes the cloud-base height slightly.
Since the minimum particle size is not sensitive to the cloud-base height ($r_{\rm coag}\propto P_{\rm b}^{1/10}$ from Equation \eqref{eq:r_coag2}), the convective adjustment has little effect on the predicted cloud-top height.

\subsection{Cloud-Top Height for Porous Cloud Particles}\label{sec:porosity}
\begin{figure*}[t]
\includegraphics[clip,width=\hsize]{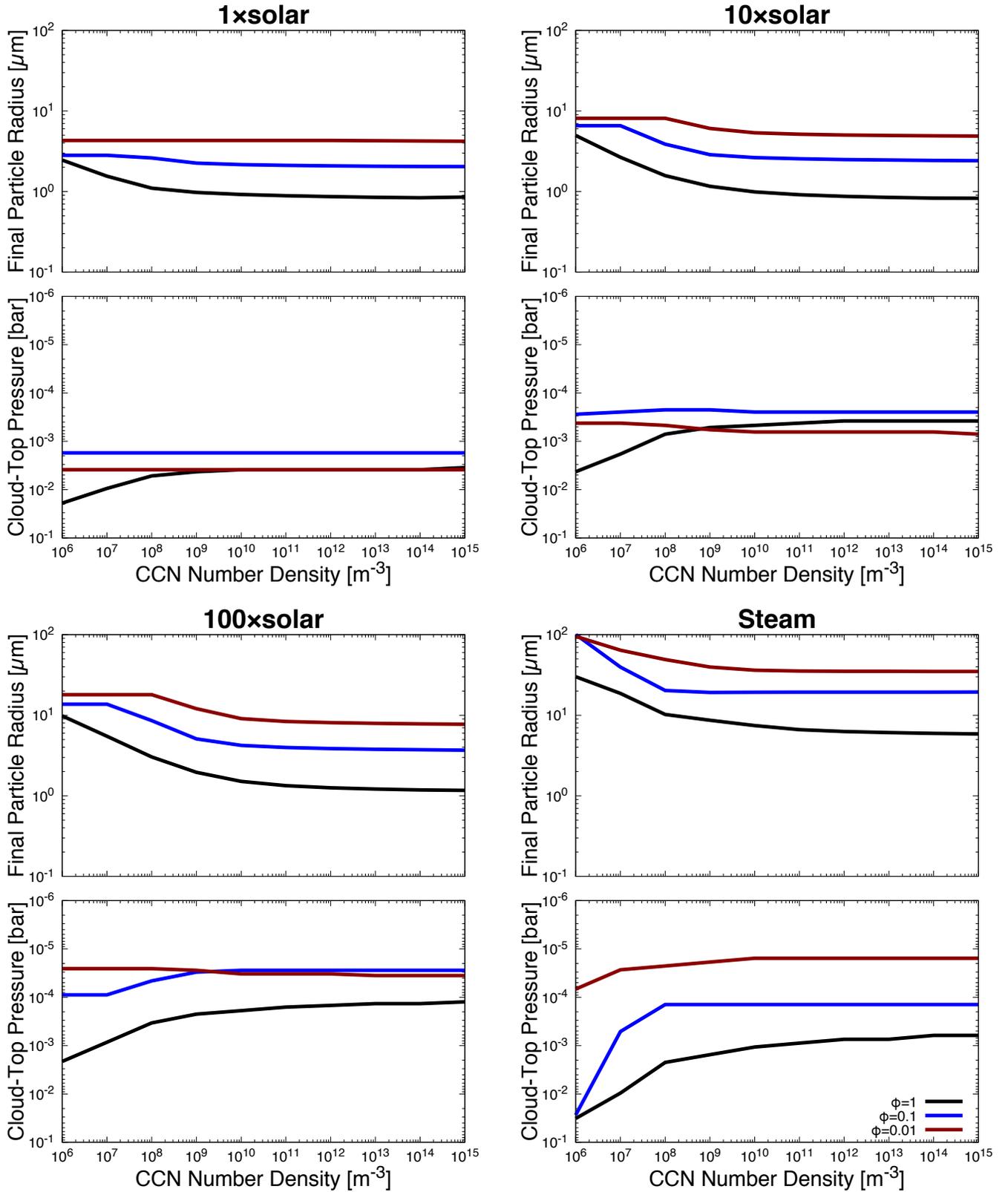}
\caption{Cloud-top pressures for various volume filling factor and atmospheric metallicity. The black, red, and blue lines denote the results for $\phi=1$, $0.1$, and $0.01$, respectively. The top and bottom panels of each block show the final particle radius and cloud-top pressure for different atmospheric metallicity. The influences of size distribution is also denoted for $\phi=0.1$ and $0.01$ using the same manner as in Figure \ref{fig:cloudtop}.
}
\label{fig:porosity}
\end{figure*}

\begin{figure}
\includegraphics[clip,width=\hsize]{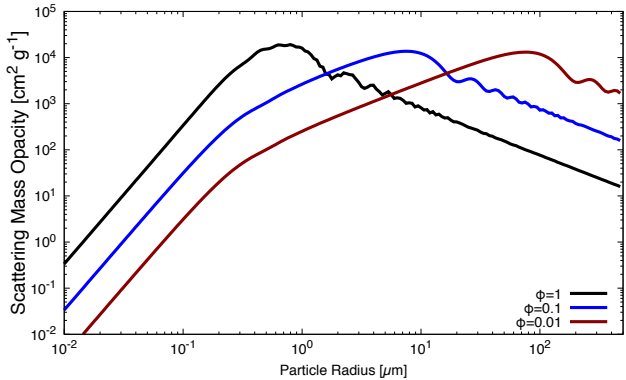}
\caption{Scattering mass opacity for porous aggregates as a function of particle size calculated by Mie theory instrumented with EMT. 
The color differences denote the differences in $\phi$. The wavelength is set as $\lambda=1.4~{\rm \mu m}$.
}
\label{fig:optical}
\end{figure}

\begin{figure*}[t]
\includegraphics[clip,width=\hsize]{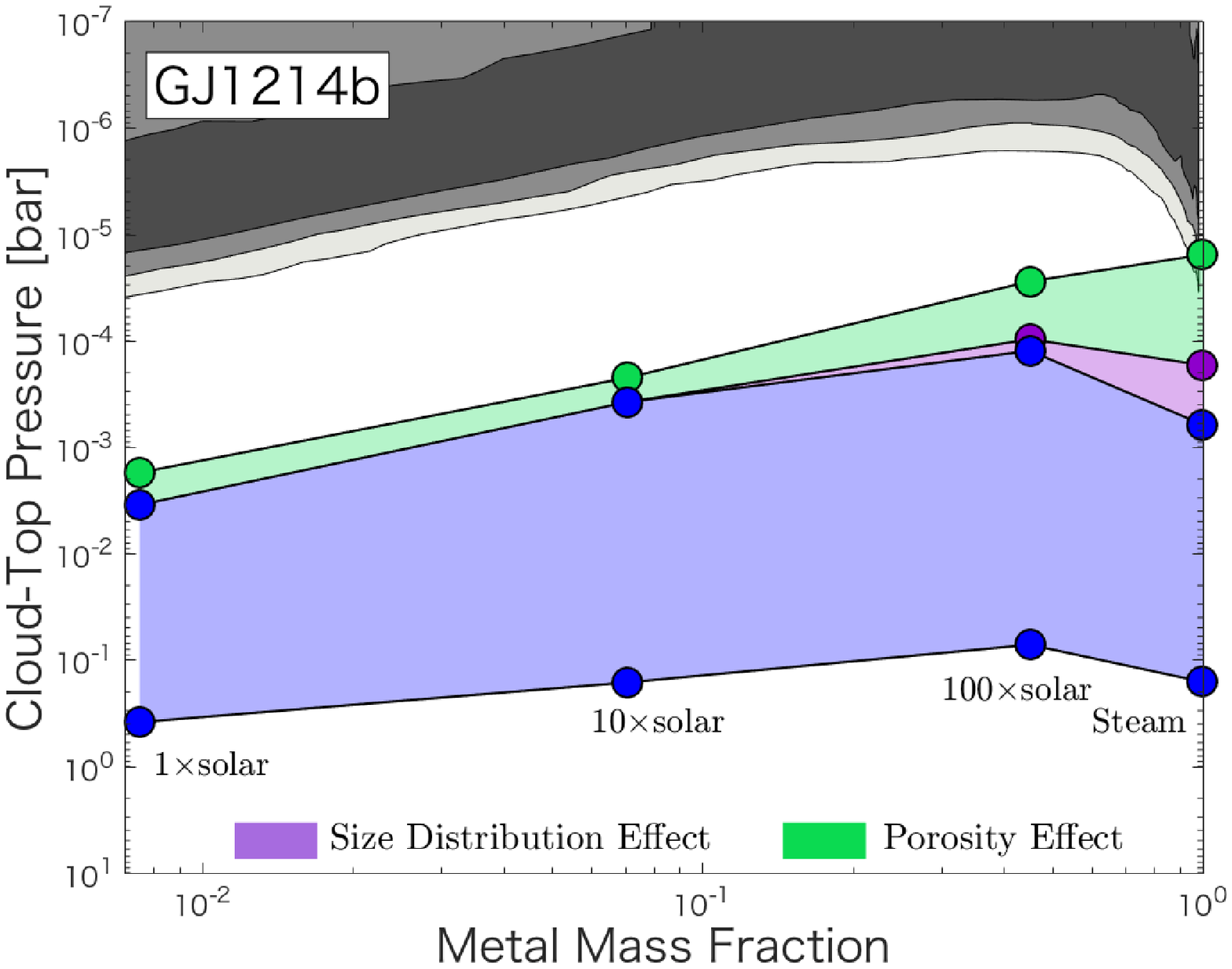}
\caption{Same as the left panel of Figure \ref{fig:GJ1214b}, but from models including the effects of size distribution (purple shaded area) and particle porosity (green shaded area).
}
\label{fig:summary}
\end{figure*}

In the calculations presented in Section \ref{sec:result} and Section \ref{sec:observation}, we assumed that cloud particles are compact and their internal density is constant.
This assumption would be valid for liquid droplets, but breaks down if solid KCl cloud particles grow into porous aggregates.
As pointed out by \citet{Marley+13}, porous aggregates are easily lofted to high altitude because they have large cross sections as compared to compact particles of the same mass.
Therefore, the predicted cloud-top height could be influenced by particle porosity.

Here we quantify the impacts of particle porosity on the predicted cloud-top height.
We introduce the volume filling factor $\phi$ defined by
\begin{equation}
\phi \equiv \frac{\rho_{\rm int}}{\rho_{\rm p}}.
\end{equation}
The volume filling factor takes $\phi=1$ for compact particles and $\phi<1$ for porous particles.
Snowflakes in the Earth are known to have $\phi=0.5$--$0.005$ \citep{Magono&Nakamura65}, while grains in protoplanetary disks could have an extremely low filling factor of $\phi \sim {10}^{-4}$ according to recent theoretical studies \citep{Okuzumi+12,Kataoka+13}.
We repeat the calculations presented in Section \ref{sec:cloudtop} by replacing $\rho_{\rm p}$ as $\rho_{\rm int}$ and varying $\phi$ from $\phi=1$ to $\phi=0.001$.

To evaluate $Q_{\rm ext}$ for porous aggregates, we calculate the effective refractive index using the effective medium theory (EMT) with the Maxwell-Garnett mixing rule \citep{Bohren&Huffman83}.
The EMT provides reasonable estimates for aggregate's absorption and scattering opacities when the particles that constitute the aggregates are smaller than the incident wavelength \citep{Voshchinnikov+07,Shen+08}.

In figure \ref{fig:porosity}, we show the final characteristic size and cloud-top height for various values of $\phi$ and atmospheric metallicities.
We find that the final characteristic size increases with decreasing $\phi$.
This is because, in the limit of high $N_{\rm CCN}$, the final characteristic size is proportional to $\phi^{-1/4}$ as indicated by Equation \eqref{eq:r_coag}.
Since $v_{\rm t}\propto \rho_{\rm p}r_{\rm c}\propto \phi^{3/4}$ in upper atmospheres (see Equation \eqref{eq:vt_eps}), porous aggregates are indeed easily lofted to high altitude as compared to compact particles.

However, the cloud-top height does not appreciably increase with decreasing $\phi$ except for the steam atmosphere (see each bottom panel of Figure \ref{fig:porosity}).
When the particle porosity is taken into account, the maximum cloud-top height is $\sim 2\times{10}^{-3}~{\rm bar}$ for the metallicity of $1\times$ solar, $\sim 2\times{10}^{-4}~{\rm bar}$ for $10\times$ solar, and $\sim 3\times{10}^{-5}~{\rm bar}$ for $100\times$ solar, respectively, which are only higher than those for compact particles by a factor of $2$--$3$.
The reason why the cloud-top height is insensitive to $\phi$ comes from the optical properties of porous aggregates, shown in Figure \ref{fig:porosity}. 
The scattering mass opacity of a porous aggregate is proportional to $\phi$ as long as $\phi<\lambda/r$, in which the aggregate itself becomes optically thin \citep{Kataoka+14}.
The two effects of reducing opacities and increasing the cloud amounts at high altitude with decreasing $\phi$ largely cancel out, explaining why the impacts of particle porosities are not drastic for hydrogen-rich atmospheres.

By contrast, for the steam atmosphere, the cloud-top height for porous aggregates can be much higher than that for compact particles.
The maximum cloud-top height is $\sim 1\times{10}^{-4}~{\rm bar}$ for $\phi=0.1$ and $\sim 2\times{10}^{-5}~{\rm bar}$ for $\phi=0.01$.
The distinct increase in cloud-top height is caused by the efficient growth via coalescence.
Because $\tau_{\rm coal}$ only depends on the cloud mass mixing ratio (see Equation \eqref{eq:t_coal}), the efficient growth via coalescence occurs even for porous aggregates.
Coalescence produces particles large enough to have a high scattering opacity, and hence the cancellation due to the effect of reducing opacities with decreasing $\phi$ does not occurs appreciably.
This is a reason why the particle porosity drastically increases the cloud-top height for the steam atmosphere.

Figure \ref{fig:summary} shows the maximum extent of KCl clouds for GJ1214 b from the models that take into account size distribution (Section \ref{sec:sizedist}) and particle porosity.
We find that the cloud-top height is still too low to be consistent with the flat spectrum of \citet{Kreidberg+14a} for hydrogen-rich atmospheres ($1\times$, $10\times$, and $100\times$ solar metallicities).
On the other hand, for the steam atmosphere with $\phi=0.01$, we find that KCl clouds can reach $\approx1.5\times {10}^{-5}~{\rm bar}$ that is equivalent to within the $3\sigma$ Bayesian credible regions of the cloud-top height ($P\approx 3\times{10}^{-5}~{\rm bar}$) reported by \citet{Kreidberg+14a}.
Since particle porosity naturally increases through coalescence, high-altitude cloud formation in the steam atmospheres, where coalescence is effective, might be a plausible explanation for the flat spectrum of GJ1214 b.

We note that the estimates for the cloud-top height given above are based on the assumption of isotropic scattering.
The forward scattering of cloud particles potentially reduces the effective cloud opacity \citep{deKok&Stam12,Robinson17} and hence produces a lower cloud top.
This effect cannot be captured here correctly because EMT tends to overestimate the degree of forward scattering of porous aggregates \citep{Shen+09,Tazaki+16,Tazaki&Tanaka18}.
The angular dependent properties of scattered light depend on the microstructure of an aggregate.
Further understanding about the microstructure and optical property of aggregates is required to verify the possibility of high-altitude cloud formation by porous aggregates.

\section{Conclusions} \label{sec:conclusions}
We have investigated how the vertical profiles of mineral clouds in super-Earths vary with the atmospheric metallicity and CCN concentration.
We used a cloud microphysical model takes into account the condensation, collision growth, and vertical transport of mineral particles in a self-consistent manner.
We have discussed how the particle size is determined by microphysical processes, and compared the predicted cloud profiles with the observations of GJ1214 b and GJ436 b.
Our main findings are summarized as follows.
\begin{enumerate}
\item The vertical profiles of mineral clouds significantly vary with CCN concentration and atmospheric metallicity.
The particle size decreases with increasing CCN concentration, and increases with increasing metallicity. 
The cloud particle's size is always larger than the minimum size determined by coagulation growth at high altitude (Equation~\eqref{eq:r_coag}).
When the mixing ratio of condensing vapor exceeds a threshold, the cloud particles grow further through coalescence.
\item Particle growth through coagulation and coalescence sets the maximum height that can be reached by a mineral cloud. 
When the mixing ratio of condensing vapor is lower than a threshold (Equation~\eqref{eq:q_cri}), the maximum cloud-top height is set by coagulation and 
 increases with increasing metallicity.
For mixing ratios above the threshold, the cloud-top height no longer increases 
with metallicity because coalescence causes further growth of the particles.
\item For GJ436 b, we have found that mineral clouds can ascend to the height suggested from the transmission spectrum \citep{Knutson+14a} for all range of metallicity ($1$--$1000\times$solar).
Since we adopted metallicity-independent eddy diffusion coefficient,
future investigation on its metallicity-dependence will allow us to determine the plausible atmospheric metallicity of GJ436 b.
\item For GJ1214 b, our model suggests that KCl clouds cannot reach the height where the presence of a cloud has been inferred from the transmission spectrum \citep{Kreidberg+14a}.
Previous cloud models suggested high-altitude clouds can form in GJ1214 b if the atmosphere's metallicity is higher than $100\times$ solar and if the cloud particle radius are around $0.5~{\rm \mu m}$ \citep{Charnay+15,Charnay+15b}. 
However, we have found that the particles always grow beyond a micron in radius through coalescence and coagulation, and suffer from ascending high enough height to explain transmission observations.
Even if the size distribution is taken into account, the height of KCl clouds is too low to be consistent with the observation of GJ1214 b because the mass-dominating particles, which is treated in our model, also dominates the total opacity in near-infrared for these particular examples.
\item Porosity evolution of cloud particles might explain the presence of the high-altitude cloud in GJ1214 b. 
We have found that KCl clouds can reach the height suggested by \citet{Kreidberg+14a} if the cloud particles have a filling factor of $0.01$ and if the atmosphere is extremely metal-enriched.
Since metal-enriched atmospheres lead to coalescence that naturally yields porous aggregates, this possibility might be a plausible solution for the flat transmission spectrum of GJ1214 b.
Our future modeling of the microstructure and optical properties of porous aggregates will pursue this possibility.

\end{enumerate}

\acknowledgments 
The authors thank the anonymous referee for insightful comments that greatly improved this paper.
We also thank {Peter} Gao and {Diana} Powell for providing paper manuscripts, and {Shigeru} Ida, {Hidekazu} Tanaka, {Xi} Zhang, {Chris} Ormel, {Ryo} Tazaki, {Yamira} Miguel, {Masahiro} Ogihara, {Yuichi} Ito, and {Yui} Kawashima for helpful comments. This work was supported by JSPS Grants-in-Aid for Scientific Research (\#15H02065, 16H04081, 16K17661) and Foundation for Promotion of Astronomy.

\appendix
\section{Evaluation of Physical Quantities}\label{appendix:1}
Here we summarize the evaluation of each physical quantity used in our calculations.

For hydrogen-rich atmosphere (metallicity of $1\times$,$10\times$,and $100\times$ solar),
we adopted the convenient formula of kinetic viscosity $\eta$, mean free path $l$, and thermal conductivity $K$ proposed by \citet{Woitke&Helling03}:
\begin{equation}\label{eq:visco}
\eta=5.877\times{10}^{-7}~{\rm Pa~s}\sqrt{T{\rm[K]}},
\end{equation}
\begin{equation}
l=1.86\times{10}^{-6}~{\rm m}\left( \frac{\rho_{\rm g}}{{10}^{-2}~{\rm kg~{m}^{-3}}}\right)^{-1},
\end{equation}
\begin{equation}
K=988\times{10}^{-5}~{\rm W~K^{-1}~{m}^{-1}}\sqrt{T{\rm [K]}}.
\end{equation}
For metal-rich cases (steam atmosphere and $1000\times$ solar),
we adopted the original formula of kinetic viscosity, i.e.,
\begin{equation}
\eta = {\displaystyle \Sigma_{\rm i}} \frac{0.499n_{\rm i}m_{\rm i}v_{\rm i}^{\rm th}}{\Sigma n_{\rm j}\pi (r_{\rm i}+r_{\rm j})^2\sqrt{1+m_{\rm i}/m_{\rm j}}},
\end{equation}
where $n_{\rm i}$, $m_{\rm i}$, $r_{\rm i}$, and $v_{\rm i}^{\rm th}=\sqrt{8k_{\rm B}T/\pi m_{\rm i}}$ are the number densities, mass, radius, and the thermal velocity of gas particles i, respectively.
In accordance with \citet{Woitke&Helling03}, we used the radii of hydrogen $r_{\rm H_{\rm 2}}=1.36~{\rm \AA}$ and $r_{\rm He}=1.09~{\rm \AA}$.
For the water, we adopted the molecular diameter used by \citet{Charnay+15}, given by
\begin{equation}
d_{\rm H_2O}=4.597~{\rm \AA}\left( \frac{T}{300~{\rm K}}\right)^{-0.3}.
\end{equation}
We also calculated the mean free path from the relation of $\eta=\rho_{\rm g}\overline{v}_{\rm th}l/3$, where $\overline{v}_{\rm th}=\sqrt{8k_{\rm B}T/\pi m_{\rm g}}$. 
In accordance with \citet{Woitke&Helling03}, we also calculated the thermal conductivity as
\begin{equation}
K=\frac{9\gamma-5}{4}\eta C_{\rm V},
\end{equation}
where $\gamma$ is the heat capacity ratio.

The diffusivity of vapor in the atmosphere is required to calculate the condensation growth.
The molecular diffusion coefficient for species i is given by \citet{Jacobson05}
\begin{equation}
D=\frac{5}{16N_{\rm A}d_{\rm i}^2\rho_{\rm g}}\sqrt{\frac{RTm_{\rm g}}{2\pi}\left( \frac{m_{\rm i}+m_{\rm g}}{m_{\rm i}} \right)},
\end{equation}
where $N_{\rm A}$ is Avogadro's number, $d_{\rm i}$ is the collision diameter.
We took the collision diameter of KCl molecules from equilibrium bond length, $d_{\rm KCl}=2.67~{\rm \AA}$ \citep{Lovas&Tiemann74}.


\end{document}